\documentclass[a4paper,11pt]{article}
\usepackage{jcappub}
\usepackage[T1]{fontenc}
\usepackage[caption=false]{subfig}
\usepackage{hyperref}
\usepackage{amsmath}
\usepackage{mathtools}
\usepackage{tablefootnote}
\usepackage{soul}
\usepackage{ulem}
\usepackage{color}
\usepackage{float}
\usepackage{multirow}
\usepackage{comment}
\usepackage[usenames,dvipsnames,svgnames,table]{xcolor}

\title{Gravitational waves $\times$ HI intensity mapping: cosmological and astrophysical applications}

\author[a,b,c]{Giulio Scelfo,}
\emailAdd{giulio.scelfo@sissa.it}
\author[d,b,c,e]{Marta Spinelli,}
\emailAdd{marta.spinelli@inaf.it}
\author[f,g,h]{Alvise Raccanelli,}
\emailAdd{alvise.raccanelli.1@unipd.it}
\author[a,b,c]{Lumen Boco,}
\emailAdd{lboco@sissa.it}
\author[a,b,c,i]{Andrea Lapi,}
\emailAdd{lapi@sissa.it}
\author[a,b,c,d]{Matteo Viel}
\emailAdd{viel@sissa.it}

\affiliation[a]{SISSA, Via Bonomea 265, 34136 Trieste, Italy}
\affiliation[b]{INFN, Sezione di Trieste, Via Bonomea 265, 34136 Trieste, Italy}
\affiliation[c]{IFPU, Institute for Fundamental Physics of the Universe, via Beirut 2, 34151, Trieste, Italy}
\affiliation[d]{INAF/OATS, Osservatorio Astronomico di Trieste, via Tiepolo 11, I-34143 Trieste, Italy}
\affiliation[e]{Department of Physics and Astronomy, University of the Western Cape, Robert Sobukhwe Road, Bellville, 7535, South Africa}
\affiliation[f]{Dipartimento di Fisica e Astronomia ``Galileo Galilei'', Universit\`a degli Studi di Padova, via Marzolo 8, I-35131, Padova, Italy}
\affiliation[g]{INFN, Sezione di Padova, via F. Marzolo 8, I-35131 Padova, Italy.}
\affiliation[h]{Theoretical Physics Department, CERN, 1 Esplanade des Particules, 1211 Geneva 23,
Switzerland}
\affiliation[i]{INAF/IRA, via Gobetti 101, 40129 Bologna, Italy}

\abstract{Two of the most rapidly growing observables in cosmology and astrophysics are gravitational waves (GW) and the neutral hydrogen (HI) distribution.
In this work, we investigate the cross-correlation between resolved gravitational wave detections and HI signal from intensity mapping (IM) experiments.
By using a tomographic approach with angular power spectra, including all projection effects, we explore possible applications of the combination of the Einstein Telescope and the SKAO intensity mapping surveys.
We focus on three main topics:
\textit{(i)} statistical inference of the observed redshift distribution of GWs;
\textit{(ii)} constraints on dynamical dark energy models as an example of cosmological studies;
\textit{(iii)} determination of the nature of the progenitors of merging binary black holes, distinguishing between primordial and astrophysical origin.
Our results show that:
\textit{(i)} the GW redshift distribution can be calibrated with good accuracy at low redshifts, without any assumptions on cosmology or astrophysics, potentially providing a way to probe astrophysical and cosmological models;
\textit{(ii)} the constrains on the dynamical dark energy parameters are competitive with IM-only experiments, in a complementary way and potentially with less systematics;
\textit{(iii)} it will be possible to detect a relatively small abundance of primordial black holes within the gravitational waves from resolved mergers.
Our results extend towards $\mathrm{GW \times IM}$ the promising field of multi-tracing cosmology and astrophysics, which has the major advantage of allowing scientific investigations in ways that would not be possible by looking at single observables separately.}

\begin{document}
\maketitle
	
\section{Introduction}\label{sec:intro}
Since the first detection of gravitational waves (GWs), originated from the merger of a Binary Black Hole (BBH) of a total mass $M_{\mathrm{tot}} \sim 60 M_\odot$~\cite{abbott:firstligodetection, abbott:firstligodetectionproperties} the interest towards the use of GWs in astrophysics and cosmology has surged, due to the possibility of studying the Universe through a new observational channel. Several detections have been made since then \cite{Abbott:O12,Abbott:O3}, opening the scientific path of gravitational waves astronomy.

Another observable that has recently emerged as extremely promising is the measurement of the integrated emission from spectral lines coming from unresolved galaxies and the diffuse intergalactic medium, the so-called Intensity Mapping (IM - see e.g.,~\cite{kovetz17:lim} for a comprehensive review). The IM technique allows probing large areas of the sky in a relatively small amount of time, since it does not aim at resolving single galaxies: it measures the intensity of a specific emission line in order to map the underlying matter distribution, treating it as a diffuse background. Since we exactly know the emission frequency of the line under study, the observed wavelength provides information on the radial position of the source, whereas its brightness temperature fluctuations describe how the underlying Large Scale Structure (LSS) is distributed.
The redshifted $21$~cm line of neutral hydrogen is one of the most promising targets for IM and several detections of the signal in cross-correlation with galaxy surveys have already been achieved (see e.g.,~\cite{Chang2010,Masui2013,Anderson2018,Wolz2021}). This tool potentially allows us to trace the LSS over a vast range of redshifts, and for the focus of this work, from the end of the reionization epoch ($z \sim 6$) to the present day \cite{Villaescusa+14:reio}. IM surveys have been proposed for the forthcoming Square Kilometre Array Observatory (SKAO) \cite{Braun2015:ska} and are ongoing on its precursor MeerKAT \cite{Santos+17,Wang+21:meerkat}, potentially bringing exquisite constraints for Cosmology \cite{SKA_redbook,maartens:ska,Santos15:SKA}. 
Other purpose-built experiments are taking data or will be build in the near future e.g. CHIME \cite{bandura2014}, FAST \cite{Hu2020},
BINGO \cite{Battye2016}, Tianlai \cite{Tianlai} and HIRAX \cite{Newburgh2016}.

Given the rapidly growing interest in both GW and IM, it is natural to investigate the synergies and the scientific output that can be obtained through their combination. In fact, cross-correlations of different tracers have been used already as a probe for cosmology. For example, the cross-correlation between e.g.,~the LSS and the Cosmic Microwave Background (see e.g.,~\cite{nolta:2004, ho:correlation, hirata:correlation, raccanelli:crosscorrelation, raccanelli:radio, raccanelli:isw, Bianchini:2014dla, Bianchini:2015fiw, Bianchini:2015yly, Mukherjee:gwxcmb}), neutrinos (see e.g.,~\cite{fang:cross}), various LSS tracers (see e.g.,~\cite{Martinez:cross,Jain:cross,Yang:cross,Paech:cross}) and even GWs (see e.g.,~\cite{Oguri:2016, raccanelli:pbhprogenitors, Scelfo18:gwxlss,Scelfo20:gws,namikawa:cross_ng,alonso:cross,Canas:sgwb,Calore:crosscorrelating,camera:gwlensing, Libanore+21, Mukherjee:gwxlss1, Mukherjee:gwxlss2,Mukherjee:sgwb,canas2021gaus}). Finally, also the IM technique has been the subject of cross-correlation studies, such as e.g.,~\cite{Schmidt13:cross,Alonso15,Kovetz16:cross,Alonso16:cross,Wolz2016:cross,Pourtsidou16:cross,Pourtsidou16:cross_2,Raccanelli16:cross,Pourtsidou17:IM,Wolz17:cross,Alonso17:lssxim,Wolz18:cross,Cunnington18:lssxim}.

The advantage of considering maps of emission line intensity as observables is not limited to the fact that they provide another LSS tracer. Intensity mapping measurements allow performing a very refined tomography: knowing the expected emission wavelength of the line under study allows for a precise and fine redshift distribution determination. In addition, IM is able to cover large cosmological volumes with respect to resolved galaxy surveys, in a relatively fast and inexpensive way.

In this work we aim at characterizing the cross-correlation signal between IM and GWs, focusing on the IM of the neutral hydrogen (HI) from the proposed 21cm IM survey with the SKAO and on resolved GW events from the merger of BBHs as detected by the Einstein Telescope (ET) \cite{Sathyaprakash:ET}.
A cross-correlation signal is expected because both HI and GWs trace the cosmic density field. Crucially, as we will see, they do so in different ways depending on some underlying assumptions on both astrophysics and cosmology.

We then present a few possible applications by studying astrophysical and cosmological tests that can be performed through the $\mathrm{GW \times IM}$ cross-correlation and forecast their potential when considering expected data from the SKAO and the ET.

Firstly, we investigate the possibility of calibrating the statistical redshift distribution of GW events thanks to the cross-correlation with IM. This idea relies on the fact that, while GWs are affected by a large redshift uncertainty (in case an electromagnetic counterpart is not available, such as for BBHs), the IM provides uniquely refined tomographic information on the observed signal. Since the HI is a good tracer of the LSS, by assuming that the BBH have astrophysical origin, we would expect them to highly cross-correlate with the LSS and, consequently, with the HI IM signal. This is a generally valid technique applicable when considering two tracers, one of which is characterized by much smaller redshift errors than the other; this was already addressed is several works in the literature (see e.g.,~references \cite{Newman_2008,Benjamin_2010,Matthews_2010,Schmidt_2013,menard2013clustering,McQuinn13:cbr,Choi_2016,Scottez_2016,Rahman_2016,Johnson_2016,Alonso17:lssxim,Daalen_2018,Cunnington18:lssxim,Alonso_2021}).
Here we investigate, to our knowledge for the first time, its potential as a method to obtain statistical redshift distributions for GW catalogs, which will provide a great improvement in dark sirens and cross-correlation studies.

Secondly, we study how this observable could help constraining cosmological models. As an example application, we focus on limits that will be possible to obtain for parameters describing the time evolution of the dark energy equation of state.

Thirdly, we tackle the issue of understanding the nature of the progenitors of the merging BBHs: evidences of the presence of Primordial Black Holes (PBHs) among the detected mergers can be found by looking at how GWs trace the underlying matter distribution (and, consequently, the HI IM signal) since different formation scenarios provide different predictions \cite{raccanelli:pbhprogenitors,Scelfo18:gwxlss}.

This manuscript is structured as follows: in section \ref{sec:formalism} we present the methodology used, introducing the mathematical formalism for the cross-correlation angular power spectra and then the Fisher matrix we use for our analyses; in section \ref{sec:tracers} we characterize our GW and HI tracers; in section \ref{sec:dNdz_reconstr} we describe the GW statistical redshift distribution calibration application; in section \ref{sec:DE} we address the dynamical DE topic; in section \ref{sec:pbhvsastro} we tackle the determination of the BBHs progenitors and in section \ref{sec:conclusions} we draw our conclusions.

\section{Methodology}\label{sec:formalism}
In this section we introduce the mathematical formalism used in this work: in section \ref{sec:formalism_Cls} we characterize the angular power spectra $C_\ell$s and then in section \ref{sec:formalism_Fisher} we describe the methodology adopted for obtaining our results, namely the Fisher matrix formalism.

\subsection{Angular power spectra}
\label{sec:formalism_Cls}
The most natural way to compute cross-correlations is by looking at the 3D angular power spectrum, $C_\ell$, therefore calculating the correlation of distributions on concentric spheres. This formalism has a long history in cosmology and was initially developed in~\cite{peebles:1973,peebles:1980}, and subsequently applied to cosmological datasets in~\cite{Regos:1989,Scharf:1992,Lahav:1993,Fisher:1994}; more recently it has been used mostly for cross-correlations (e.g.,~\cite{nolta:2004}). The advantage of this formalism resides in the fact that it naturally includes effects coming from large angular separations, the curvature of the sky, and that it makes use of directly observable quantities such as angles and redshifts. The drawback of having to calculate a large number or auto- and cross- bin correlations in the case of many narrow redshift bins does not apply here as we do not have very good radial information for the GW maps. Moreover, recent theoretical developments allow us to compute a large number of correlations in very short times (see~\cite{gebhardt:2018fast, assassi:2017}).

In the following we describe the general formalism for this calculation for resolved events, such as in the cases of e.g.,~galaxies or GW events, where $C_\ell$s indicate \textit{number counts} angular power spectra. In section \ref{sec:tracer_HI} we describe how this formalism can easily be extended to non-resolved tracers such as HI from intensity mapping.

Defining the number count fluctuations of a tracer $X$ at redshift $z$ and direction $\hat{n}$ as $\delta^{X}(z, \hat{n})$, we can expand it in spherical harmonics $Y_{\ell m}(\hat{n})$ using the harmonic coefficients $a_{\ell m}^{X}(z)$, as
\begin{equation}
\delta^{X}(z, \hat{n})=\sum_{\ell m} a_{\ell m}^{X}(z) Y_{\ell m}(\hat{n}).
\end{equation}
The relation between the harmonic coefficients and the \textit{observed} angular power spectrum $\tilde{C}^{XY}_\ell(z_i,z_j)$ (describing the cross-correlation of tracer $X$ in redshift bin $z_i$ with tracer $Y$ in bin $z_j$) is the covariance of the coefficients of the spherical harmonics expansion, given by:
\begin{equation}
\label{eq:clsxy}
\langle a^X_{\ell m}(z_i) a^{Y^*}_{\ell'm'}(z_j)\rangle = \delta_{\ell \ell'} \delta_{mm'}{\tilde{C}}^{XY}_\ell(z_i,z_j), 
\end{equation}
where $\delta$ stands for the Kronecker delta. Note that this expression is strictly valid only in the case of isotropic and homogeneous fields (see~\cite{RV1}). For the case of this paper, however, we will continue using the standard formalism derived from equation \eqref{eq:clsxy} because of the poor angular and radial resolution of GW maps.

The $a_{\ell m}^X(z_i)$ coefficients are built from the partial wave coefficients of the signal and of the noise
\begin{equation}
a_{\ell m}^X(z_i) = s_{\ell m}^X(z_i) + n^X_{\ell m}(z_i).
\end{equation}
The {\it observed} angular power spectrum is then written as
\begin{equation}
\tilde{C}^{XY}_\ell(z_i,z_j) = C^{XY}_\ell(z_i,z_j) + \mathcal{N}^{XY}_\ell(z_i,z_j).
\label{eq:tilde_Cls}
\end{equation}
The cross-correlation angular power spectrum of the signal and noise are computed from the signal wave coefficients as (see e.g.~\cite{peebles:1973,Regos:1989,Scharf:1992,Lahav:1993,Fisher:1994,challinor:deltag,bonvin:cl})
\begin{align}
    \langle s^X_{\ell m}(z_i) s^{Y^*}_{\ell' m'}(z_j) \rangle = \delta_{\ell\ell'} \delta_{mm'} C^{XY}_\ell(z_i,z_j) \\
    \langle n^X_{\ell m}(z_i) n^{Y^*}_{\ell' m'}(z_j) \rangle = \mathcal{N}^{XY}_{\ell m}(z_i,z_j).
\end{align}
For the GW detecion we set up a SNR of 8 to guarantee detections and include a shot noise term, while we use a combination of instrumental and foreground noises for IM (see section \ref{sec:tracers} for details on their explicit expressions).
We also assume that signal and noise are statistically independent,
$
\langle s^X_{\ell m}(z_i) n^{Y^*}_{\ell' m'}(z_j) \rangle = 0.
$
The 3D angular power spectrum of tracers $\{X,Y\}$ at redshifts $\{z_i,z_j\}$ can be written as
\begin{equation}
C_{\ell}^{XY}(z_i,z_j) = \frac{2}{\pi} \int\frac{dk}{k} \mathcal{P}(k) \Delta^{X,z_i}_{\ell}(k) \Delta^{Y,z_j}_{\ell}(k),
\label{eq:Cls}
\end{equation}
where $\mathcal{P}(k)= k^3P(k)$ is the primordial  power spectrum and
\begin{equation}
\Delta^{X,z_i}_{\ell}(k) = \int_{0}^{\infty} dz \frac{dN_X}{dz}W(z,z_i,\Delta z_i)\Delta^X_\ell(k,z).
\label{eq:Delta_l}
\end{equation}
Here $W(z,z_i,\Delta z_i)$ are observational window functions related to the specific experiment centered at $z_i$ with half-width $\Delta z_i$ and $\dfrac{dN_X}{dz}$ stands for the source number density per redshift interval. Note that the integral of $\displaystyle W(z,z_i,\Delta z_i)\frac{dN_X}{dz}$ is normalized to unity. Finally, $\Delta^X_\ell(k,z)$ is the angular number count fluctuation of the $X$ tracer, which is the sum of density ($\mathrm{den}$), velocity ($\mathrm{vel}$), lensing ($\mathrm{len}$) and gravity ($\mathrm{gr}$) effects~\cite{bonvin:cl, challinor:deltag}:
\begin{equation}
\Delta_\ell^{X}(k,z) = \Delta^\mathrm{den}_{\ell}(k,z) + \Delta^\mathrm{vel}_\ell(k,z) + \Delta^\mathrm{len}_\ell(k,z) + \Delta^\mathrm{gr}_\ell(k,z).
\label{eq:numbercount_fluctuation}
\end{equation}
The reader interested in the full expression of the terms in equation \eqref{eq:numbercount_fluctuation} can find them in appendix \ref{app:deltas}.
In this work we computed angular power spectra using \texttt{Multi\_CLASS}\footnote{Publicly available at \url{https://github.com/nbellomo/Multi_CLASS}.}, the modified version of \texttt{CLASS} \cite{blas:class,didio:classgal} presented in~\cite{Bellomo20:multiclass, Bernal20:multiclass} which allows to compute cross-correlations between different tracers.

In the following, we list and briefly describe the relevant physical quantities for the computation. More details about the specifics for the tracers considered in this work (redshift distributions, biases, redshift binning, etc.) are provided in sections \ref{sec:tracer_GW} and \ref{sec:tracer_HI}.
\begin{itemize}
    \item \textit{Redshift distribution} $\dfrac{dN_X}{dz}$: observed source number density per redshift interval of tracer $X$. It appears in equation \eqref{eq:Delta_l} for the $C_\ell$s computation, in which its shape is the only significant feature, whereas its overall amplitude has no relevance due to the normalization. The total number of objects as a function of redshift is instead used when computing the shot noise of resolved sources. We have checked that using a top-hat or a gaussian window function does not significantly affect our findings.
    \item \textit{Bias} $b_X$: it describes the relation between a given observable $X$ and the underlying distribution of matter that it traces (see e.g.,~\cite{Kaiser:bias, Bardeen:bias, Mo:smallhalosbias, matarrese:clusteringevolution, dekel:stochasticbiasing, benson:galaxybias, peacock:halooccupation, Desjacques:bias}). Considering the linear bias formulation and indicating the local contrasts of matter and tracer $X$ at position $x$ respectively by $\delta(x)$ and $\delta_X(x)$, the bias is defined as $\delta_{X}(x) \equiv \frac{n_{X}(x)-\bar{n}_{X}}{\bar{n}_{X}}=b_{X} \delta(x)$, where $n_{X}$ is the comoving density of tracer X and $\bar{n}_{X}$ is its mean value. This physical quantity is a linear factor in the density term of equation \eqref{eq:numbercount_fluctuation}. In this work we will also make use of the quantity $\bar{b}_{X}$, which provides an average value of the bias of tracer X in a considered redshift range $[z_{\rm min},z_{\rm max}]$, weighted for the tracer redshift evolution:
    \begin{equation}
    \bar{b}_X = \frac{{\displaystyle \int_{z_\mathrm{min}}^{z_\mathrm{max}} dz\ b_{X}(z) \frac{d^2N_X}{dz d\Omega}}}{{\displaystyle \int_{z_\mathrm{min}}^{z_\mathrm{max}} dz \frac{d^2N_X}{dz d\Omega}}}
    \label{eq:bias_def}
    \end{equation}
    with $X = \{\mathrm{GW,IM}\}$. In this way, estimating a mean value for this parameter, we take into account which redshift interval (i.e. bias values) weights the most.
    \item \textit{Magnification bias} $s_X(z)$: it quantifies how the observed surface density of sources of tracer X is influenced by gravitational lensing effects~\cite{turner:magnificationbias}. The observable result is given by the contribution of two opposite effects: whereas the number of observed sources can grow thanks to a magnification effect of the incoming flux, an increase of the area lowers the observed number density of objects. The magnification bias is a dominant term in the lensing contribution of equation \eqref{eq:numbercount_fluctuation}, but affects also the velocity and gravity terms.
    \item \textit{Evolution bias} $f_X^{\mathrm{evo}}$: this term is present due to the fact that the {\it absolute} number of objects of a tracer X may not be conserved over time due to the possible formation of new objects. It reads as~\cite{challinor:evolutionbias, jeong:evolutionbias, bertacca:evolutionbias}:  $f^\mathrm{evo}_\mathrm{X}(z) = \frac{d\ln\left(a^3\frac{d^2N_\mathrm{X}}{dzd\Omega}\right)}{d\ln a}$, where $a$ is the scale factor and $\frac{d^2N_\mathrm{X}}{dzd\Omega}$ is the absolute distribution of objects of tracer $X$, which in principle is not the same as the observed one introduced above. The evolution bias appears only in sub-leading contributions to equation \eqref{eq:numbercount_fluctuation}.
\end{itemize}

We have already mentioned that the bias parameters introduced above come into play in quantifying the angular number count fluctuations of equation \eqref{eq:numbercount_fluctuation} (see appendix \ref{app:deltas} for full expressions). In the following we explicitly summarize the dependence of each of the number counts contributions on the bias parameters ($b_X$, $s_X$, $f_X^{\mathrm{evo}}$):
\begin{equation}\label{eq:biases_and_deltas}
\begin{cases}
\Delta_\ell^\mathrm{den} = \Delta_\ell^\mathrm{den}(b_X) \\
\Delta_\ell^\mathrm{vel} = \Delta_\ell^\mathrm{vel}(s_X, f_X^{\mathrm{evo}}) \\ 
\Delta_\ell^\mathrm{len} = \Delta_\ell^\mathrm{len}(s_X) \\
\Delta_\ell^\mathrm{gr} = \Delta_\ell^\mathrm{gr}(s_X, f_X^{\mathrm{evo}})
\end{cases} 
\end{equation}
where dependencies on $k$ and $z$ are implied.

\subsection{Fisher analysis}
\label{sec:formalism_Fisher}
In this work we make use of the Fisher analysis methodology, which we briefly sketch here to introduce the general formalism we adopt.
Assuming GWs and IM signals as the two tracers, we divide the total redshift interval surveyed by considered GW experiments in $N_{\mathrm{bins}}^{\mathrm{GW}}$ bins, with amplitude $\Delta z^{\mathrm{GW}}$, and the signal from intensity mapping distributed among $N_{\mathrm{bins}}^{\mathrm{IM}}$ redshift bins with amplitude $\Delta z^{\mathrm{IM}}$.

Considering the observed power spectra $\tilde{C}_\ell$s and a generic set of parameters $\{\theta_n\}$ for the Fisher analysis, we can organize our data in the (symmetric) matrix $\mathcal{C}_\ell$ as
\begin{equation}
\mathcal{C}_\ell=
\begin{bmatrix}
\tilde{C_\ell}^{\mathrm{IM\,IM}}(z_1^{\mathrm{IM}},z_1^{\mathrm{IM}}) & ... & \tilde{C_\ell}^{\mathrm{IM\,IM}}(z_1^{\mathrm{IM}},z_N^{\mathrm{IM}}) & \tilde{C_\ell}^{\mathrm{IM\,GW}}(z_1^{\mathrm{IM}},z_1^{\mathrm{GW}}) & ... &
\tilde{C_\ell}^{\mathrm{IM\,GW}}(z_1^{\mathrm{IM}},z_N^{\mathrm{GW}}) \\
& ... & \tilde{C_\ell}^{\mathrm{IM\,IM}}(z_2^{\mathrm{IM}},z_N^{\mathrm{IM}}) & \tilde{C_\ell}^{\mathrm{IM\,GW}}(z_2^{\mathrm{IM}},z_1^{\mathrm{GW}}) & ... &
\tilde{C_\ell}^{\mathrm{IM\,GW}}(z_2^{\mathrm{IM}},z_N^{\mathrm{GW}})\\
& ... & $\vdots$ & $\vdots$ & ... & $\vdots$\\
& & \tilde{C_\ell}^{\mathrm{IM\,IM}}(z_N^{\mathrm{IM}},z_N^{\mathrm{IM}}) & \tilde{C_\ell}^{\mathrm{IM\,GW}}(z_N^{\mathrm{IM}},z_1^{\mathrm{GW}})  & ... &
\tilde{C_\ell}^{\mathrm{IM\,GW}}(z_N^{\mathrm{IM}},z_N^{\mathrm{GW}}) \\
&  &  & \tilde{C_\ell}^{\mathrm{GW\,GW}}(z_1^{\mathrm{GW}},z_1^{\mathrm{GW}}) & ... &
\tilde{C_\ell}^{\mathrm{GW\,GW}}(z_1^{\mathrm{GW}},z_N^{\mathrm{GW}})\\
& & & & ... & $\vdots$ \\
&  & & & & \tilde{C_\ell}^{\mathrm{GW\,GW}}(z_N^{\mathrm{GW}},z_N^{\mathrm{GW}})\\
\end{bmatrix},
\label{eq:C_matrix_multi}
\end{equation}
The $\mathcal{C}_\ell$ matrix has dimensions of $(N_{\mathrm{bins}}^{\mathrm{IM}}+N_{\mathrm{bins}}^{\mathrm{GW}}) \times (N_{\mathrm{bins}}^{\mathrm{IM}}+N_{\mathrm{bins}}^{\mathrm{GW}})$. We remind the reader that the tilde symbol stands for \textit{observed} $C_\ell$s.

The $\mathcal{C}_\ell$ matrix is then used to compute the Fisher matrix elements as
\begin{equation}\label{eq:Fisher_matrix}
F_{\alpha \beta}=f_\mathrm{sky}\sum_\ell \frac{2\ell+1}{2} \mathrm{Tr}\left[\mathcal{C}_\ell^{-1} (\partial_\alpha \mathcal{C}_\ell)\mathcal{C}_\ell^{-1}(\partial_\beta \mathcal{C}_\ell)\right],
\end{equation}
where $\partial_{\alpha}$ indicates the partial derivative with respect to the parameter $\theta_{\alpha}$ and $f_\mathrm{sky}$ is the fraction of the sky covered by the intersection of IM and GW surveys. The sum over multipoles $\ell$ is performed up to a maximum value $\ell_{\rm max}$, which corresponds to the achievable angular resolution for the considered sources and instruments. For GW events an accurate estimate of $\ell_{\rm max}$ would depend on (not limited to) redshift, SNR of the events and other source properties such as mass and spin. Since a rigorous analysis for the estimation of this parameter goes beyond the scope of this work, we use the constant threshold of $\ell_{\rm max}=100$, which provides a general plausible value for ET (as also performed in other studies, see e.g., \cite{raccanelli:pbhprogenitors,Scelfo18:gwxlss,Scelfo20:gws}). All scales smaller than it are conservatively cut from the analysis.

Finally, the Fisher-estimated marginal error on the parameter $\theta_{\alpha}$ is given by $\sqrt{(F^{-1})_{\alpha \alpha}}$. According to statistics and estimation theory, the so called Cram\'er-Rao bound provides the smallest error that one should expect to achieve in reality: errors on parameters deriving from ``real-life'' experiments are expected to be equal or higher than the Fisher estimated errors (Cram\'er-Rao inequality), where the equality stands only in the case of gaussian likelihood. Even though this is often an approximation and the Fisher approach may not always give precise results, it still remains an easy and quick method to provide forecasts for planned experiments.

\section{Observables}
\label{sec:tracers}
In this section we characterize the considered tracers: resolved GWs from BBH mergers and the HI signal from intensity mapping experiments. In table \ref{tab:tracers} we summarize the redshift distributions for our tracers, as mentioned in sections \ref{sec:tracer_GW} and \ref{sec:tracer_HI}.

\begin{table}[]
\centering
\begin{tabular}{|c|c|c|}
\hline
\textbf{\:\:\:\: Tracer \:\:\:\:}         & \:\: \textbf{GW (ET)} \:\:           & \:\: \textbf{IM (SKAO)} \:\:                \\ \hline
z range        & \multicolumn{2}{c|}{{[}0.5-3.5{]} \:\:\:} \\ \hline
$N_{\rm bins}$ & 3             & 30                 \\ \hline
$\Delta z$     & 1.0           & 0.1                \\ \hline
\end{tabular}
\caption{Chosen redshift specifics and experiments for the two tracers considered in this work. Note that the redshift ranges do not necessarily correspond to the best achievable from the indicated surveys.}
\label{tab:tracers}
\end{table}

\subsection{Gravitational Waves}\label{sec:tracer_GW}
As first tracer we consider GWs from resolved mergers of BBHs, as detected by the Einstein Telescope (ET) experiment, as currently planned in~\cite{Sathyaprakash:ET}. We study this tracer for $N_{\mathrm{bins}}^{\mathrm{GW}}=3$ redshift bins with width $\Delta z^{\mathrm{GW}}=1.0$ in the redshift range $[0.5-3.5]$. Even though the ET instrument would be able to detect BBH mergers outside this redshift range, we limit our analysis to it because it is the most optimal redshift range for the SKAO-Mid band IM survey (SKAO-MID) \cite{SKA_redbook} (i.e.,~the survey we consider for our HI tracer). Considering GWs events beyond this limit would not help our analysis because we would not have any HI signal to cross-correlate them with.

It is worth noting that the ET is not the only planned third-generation GW detector. Another promising experiment is given by the Cosmic Explorer (CE) \cite{CE:2019}. We anticipate here that, within our framework, results for ET and CE are quite similar, although slightly optimistic for the latter. We thus concentrate on the more conservative ET for the rest of the paper.

Note that we consider such large bins in order to take into account any possible luminosity distance uncertainty on the observed GWs events, also maintaining an approach as independent as possible on cosmological parameters. Indeed, the bin width $\Delta_z^{\rm GW}=1.0$ is larger than any redshift uncertainty estimated for LIGO/Virgo sources (see e.g., Table 6 of reference \cite{Abbott:O3}), which will be even smaller for third-generation observatories. As a matter of fact, measurements of BBH mergers are associated to an uncertainty on the luminosity distance, which can be connected to a redshift uncertainty only by assuming a specific cosmology. An error on the assumed cosmology leads to a wrong assumption on the redshift of the event (and on its error). Since we are making use of a statistical tomographic approach, the main important element here is that the containing redshift bin for each observed event is the appropriate one. Assuming large bins for GWs makes this assumption safer, i.e.,~even when making errors on the assumed cosmology, the event-bin mapping would not be biased for most of the events. On the other hand, assuming smaller bins for GWs might provide more information and more optimistic forecasts, but it could lead to biased results if the wrong cosmology is assumed when actually performing these applications with real future data We have tested the impact on our results on the bin width choice. As way of example, we found that by reducing $\Delta_z^{\rm GW}$ from 1.0 to 0.5 (doubling the number of GW bins) the forecasts on the bias parameter $\bar{b}_{\rm GW}$ are $40\%$ more optimistic, since a more refined tomographic information is being exploited. Nonetheless, this shows that results could be even more promising than those reported in this paper, but we preferred to choose a more conservative approach by adopting larger GW redshift bins, safely getting rid of any possible bias due to redshift or cosmology related errors. This makes the forecasts presented in this manuscript almost ``cosmology-agnostic'' and independent on reasonable errors on the single sources redshifts.

\begin{figure}
	\centering
	\includegraphics[width=1.05\linewidth]{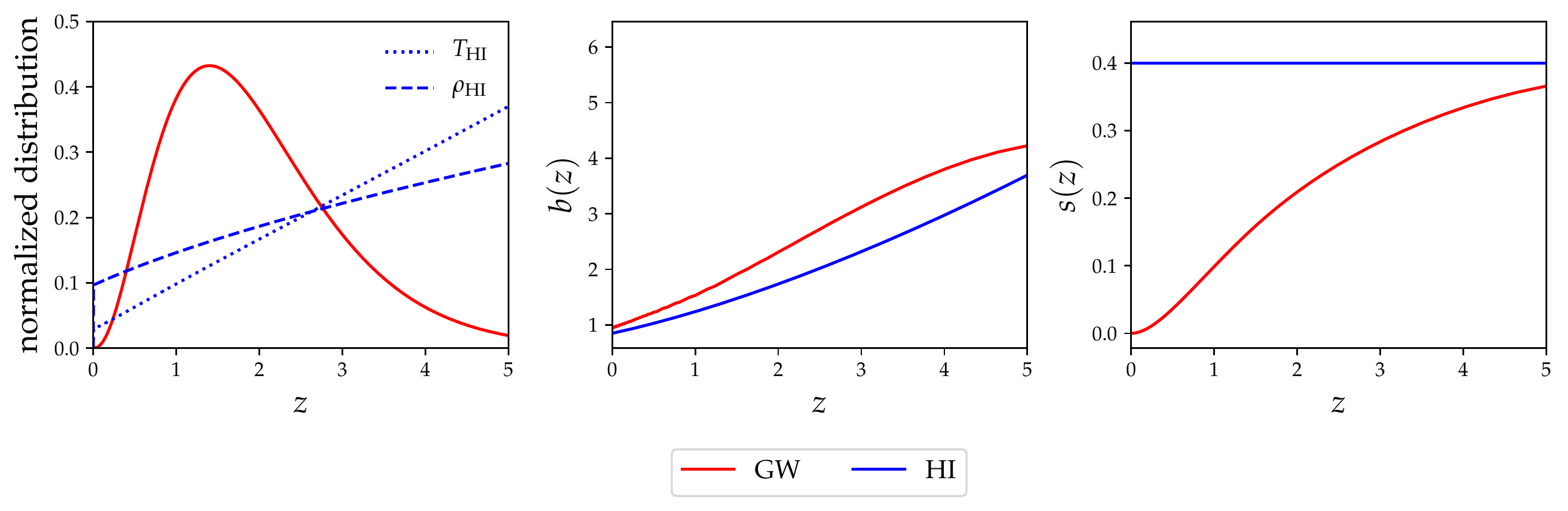}
	\caption{Specifics for the astrophysical GW and HI tracers considered in this work. \textit{Left:} normalized redshfit dependence ($dN/dz$ for GWs; $T_{\rm b}(z)$ and $\rho_{\rm HI}(z)$ for HI). \textit{Center:} bias $b(z)$. \textit{Right:} magnification bias $s(z)$.}
	\label{fig:tracers}
\end{figure}

We characterize this tracer following prescriptions from references \cite{Boco19:gws,Scelfo20:gws}, that we summarize as:
\begin{itemize}
    \item \textit{Redshift distribution:} the redshift distribution of GWs events from BBHs mergers detected by the ET in the considered redshift range is taken from reference \cite{Boco19:gws} and can be analytically interpolated in our redshift range as:
    \begin{equation}\label{eq:dNdz_GWs}
        \dfrac{dN_{\rm GW}}{dz} = A z^{b} \exp({-c z})
    \end{equation}
    with $A=0.825\cdot 10^5$, $b=2.40$, $c=1.71$, assuming an observation time of $T_{\rm obs}^{\rm GW}=1$ yr and $f_{\rm sky}=1$. Integrating this function with these specifics in our considered redshift range $0.5<z<3.5$ provides a total of $\sim 3.5 \cdot 10^{4}$ BBH mergers detections. Changes in $T_{\rm obs}^{\rm GW}$ and $f_{\rm sky}$ act simply as a re-scaling of the overall amplitude. The expression in equation \eqref{eq:dNdz_GWs} is obtained interpolating the results of the semi-analytical treatment of \cite{Boco19:gws}, in which galactic star formation rate functions, dependence of compact remnant masses on metallicity, time delays and stellar and binary evolution prescriptions are taken into account. Among astrophysical uncertainties, their findings are in agreement with other studies based on combining population synthesis simulations (e.g.,~\cite{dominik+13,dominik+15,demink+13,spera+15,spera+17,spera+19,giacobbo+18}) with recipes on the cosmic Star Formation Rate (SFR) density and metallicity distributions inferred from observations (e.g.,~\cite{belczynski:massivebhsmergers,lamberts+16,cao+18,elbert+18,li+18,neijssel+19}).
    Finally, note that the BBH merger rate is normalized to a local value of $30\: \rm Gpc^{-3} yr^{-1}$ at $z=0$, in agreement with observed data from the first half of the LIGO/Virgo collaboration O3 run \cite{Abbott:O3run}. The normalized redshift distribution for GW events from astrophysical BBH mergers is provided in the left panel of figure \ref{fig:tracers}.
    \item \textit{Bias:} we make use of the findings of reference \cite{Scelfo20:gws}, in which the bias of GWs from BBH mergers is determined through an abundance matching technique (see e.g.,~\cite{aversa+15}), associating the luminosity/SFR of the host galaxy to the mass of the hosting dark matter halo and then matching to a galaxy with given SFR the bias of the associated halo. Finally, characterizing BBH mergers with the same bias of their host galaxies, the final bias expression is obtained by taking into account which galaxy types proportionately contribute most to the detected merger rate. This quantity is provided in the central panel of figure \ref{fig:tracers} and can be interpolated up to $z \sim 3.5$ as:
    \begin{equation}\label{eq:bias_GWs}
        b_{\rm GW}(z) = a \exp(b z^d)+z^c
    \end{equation}
    with $a=0.948$, $b=-0.553$, $c=0.996$, $d=1.034$. The authors of reference \cite{Libanore+21} find a general agreement to the behaviour of these prescriptions, up to the redshift range considered in this work. Note that this determination of the bias, as well as the $dN/dz$, may also depend on other important astrophysical parameters, such as IMF and metallicity, and on the considered merging channel, such as isolated binaries, dynamical merger in stellar and nuclear stars clusters, mergers of PopIII stars, etc. References \cite{Boco19:gws,Scelfo20:gws}, on which our assumptions are based, do not take into account merging efficiency dependencies on metallicity or IMF variations, and consider an isolated binaries merging channel. A more technical treatment of these quantities is beyond the scope of this work.
    \item \textit{Magnification bias:} this quantity is defined as the logarithmic slope of the redshift distribution of detected events computed at detectability limit $\rho=\bar\rho$:
\begin{equation}
    s_{\rm GW}(z)=-\left.\frac{  d\log_{10}{\left(\frac{  d^2{N}_{\rm GW}(z,>\rho)}{  dz\,d\Omega}\right)}}{  d\rho}\right|_{\rho=\bar\rho}\, ,
\end{equation}
where $\rho$ is the Signal-to-Noise ratio for each GWs event. Usually, detection of a GWs signal is considered solid for $\rho>\bar{\rho}=8$. The magnification bias obtained from \cite{Scelfo20:gws}, and used in this work, is plotted in the right panel of figure \ref{fig:tracers}.
\item \textit{Evolution bias:} by definition, the evolution bias for GWs events is straightforwardly given by
\begin{equation}
f^\mathrm{evo}_\mathrm{GW}(z) = \frac{d\ln\left(a^3\frac{d^2{N}_{\rm GW}(z)}{dzd\Omega}\right)}{d\ln a}.    
\end{equation}
\end{itemize}
In figure \ref{fig:tracers} we provide redshift distribution and biases values for our GW tracer of astrophysical origin, together with the same quantities characterizing the HI from IM.
\section*{Noise sources} 
We characterize the considered GWs events with a shot-noise component as the only noise source to the angular power spectra:
\begin{equation}\label{eq:shotnoise_GWs}
C_{\ell}^{\mathrm{N, GW}}(z_i,z_j) = C_{\ell}^{\mathrm{shot}}(z_i,z_j) = \dfrac{\delta_{ij}}{\bar{n}(z_i)}
\end{equation}
where $\delta_{ij}$ is the Kronecker delta and $\bar{n}(z_i)$ is the mean number density of sources in the $i^{th}$ redshift bin. It affects the $C_\ell(z_i,z_i)^{\mathrm{GW,GW}}$ entries (i.e. only cross-correlations between the same GW tracer among the same redshift bin).

\subsection{HI Intensity Mapping}\label{sec:tracer_HI}
In this section we characterize our second tracer: the forecasted measure of the HI distribution by a SKAO Mid-band (SKAO-MID) intensity mapping survey \cite{SKA_redbook, Dewdney:ska, maartens:ska}. In the following sections we describe how the resolved sources formalism for the angular power spectra can be easily translated to be applied to the unresolved HI IM case, and we characterize the specifics for this observable.

Throughout this work we consider the HI tracer in the redshift range $[0.5-3.5]$, divided in bins of width $\Delta z^{\mathrm{IM}}=0.1$, for a total of $N_{\mathrm{bins}}^{\mathrm{IM}}=30$ redshift bins. This is expected to be around the optimal redshift range for the SKAO-MID survey \cite{SKA_redbook}.

\subsubsection{From resolved sources to IM}
As provided in section \ref{sec:formalism_Cls}, the cross-correlation formalism for resolved sources of different tracers is well defined and comprehensive of relativistic effects. When working with the IM signal of any line (the HI in our case), we do not deal with number counts, since we are looking at the ensemble of unresolved sources. However, the unresolved tracer case can be treated adapting the same formalism (e.g., \cite{Alonso15,Hall13}). The following points are specifically referred to the HI, but are valid for any other line. In particular:
\begin{itemize}
	\item \textit{Redshift distribution:} while we characterize resolved sources with a redshift distribution of their number counts $dN_{ X}/dz$, for the HI intensity mapping case we shall consider the HI comoving density distribution defined in \cite{Crighton15:HI} as $\rho_{\mathrm{HI}}(z)=\Omega_{\rm HI}(z)\rho_{\rm crit, 0}$ and the mean brightness temperature $T_b(z)$. Their explicit expressions are (see e.g., \cite{Crighton15:HI,Battye:T_HI}):
	\begin{equation}
	\rho_{\rm HI}(z)=4(1+z)^{0.6} 10^{-4}\cdot \rho_{\rm crit, 0}
	\end{equation}
	\begin{equation}
	T_{b}(z)=44 \mu \mathrm{K}\left(\frac{\Omega_{\mathrm{HI}}(z) h}{2.45 \times 10^{-4}}\right) \frac{(1+z)^{2}}{E(z)},
	\end{equation}
where $\rho_{\rm crit, 0}$ is the critical density today and $E(z) = H(z)/H_0$. Since the density $\rho_{\rm HI}$ provides the redshift dependence of the \textit{absolute} redshift distribution of HI atoms, whereas the mean brightness temperature $T_{b}$ is a directly \textit{observed} physical quantity through IM, we make use of $T_{b}$ in place of the observed redshift distribution of equation \eqref{eq:Delta_l} and of $\rho_{\rm HI}$ to compute the evolution bias term $f_{\rm evo}(z)$. Both $\rho_{\rm HI}(z)$ and $T_{b}$ redshift dependencies are plotted in the left panel of figure \ref{fig:tracers}.

\item \textit{Bias:} we can treat the bias analogously as it is done for the resolved sources case. In this work we use the following analytic expression obtained fitting results from reference \cite{Spinelli20:HI}:
	\begin{equation}
	b_{\rm HI}(z) = a(1+z)^{b}+c,
	\end{equation}
	with $a=0.22$, $b=1.47$ and $c=0.63$. This quantity is plotted in the central panel of figure \ref{fig:tracers}.
	This prescription originates from the outputs of a semi-analytical model for galaxy formation that include an explicit treatment of neutral hydrogen and are in agreement with the findings of \cite{Villaescusa+18:ingr} based on Illustris TNG hydro-dynamical simulations. The bias is expected to be around unity at low $z$ (e.g. $\sim 0.85$ at $z \sim 0.06$ \cite{Martin_2010}), where the HI is strongly present in young galaxies with high Star Formation Rates  \cite{Anderson2018}.
	In order to make sure that uncertainties on the HI bias at higher $z$ do not affect the conclusions of our work, we have checked that performing the same analysis with the extreme hypothesis of a constant unitary value of $b_{\rm HI}(z)$ leads to a change on the Fisher estimated errors for cosmological parameters below $15\%$.
\item \textit{Magnification bias:} when treating any IM experiment, the magnification bias assumes the value
	\begin{equation}
	s_{\rm HI}(z)=0.4,
	\end{equation}
	which corresponds to the absence of lensing effects. This is due to the fact that the observed physical quantity is a surface brightness (instead of number counts) which is not altered by this type of phenomena (see e.g., \cite{Hall13} and references therein).
	\item \textit{Evolution bias:} as mentioned above, this quantity is obtained analogously to the resolved sources case, substituting the redshift distribution for resolved sources with the density distribution:
	\begin{equation}
	f_{\rm HI}^{\rm evo}(z)=\dfrac{d\ln \rho_{\rm HI}(z)}{d\ln a}
	\end{equation}
	with $a$ being the scale factor.
\end{itemize}

\subsubsection{Noise sources}
When considering $C_\ell$s including the IM component (both $\mathrm{IM \times IM}$ and $\mathrm{GW \times IM}$ cases) we express the relation between theoretical $C_\ell^{XY}$ (computed with \texttt{Multi\_CLASS}) and the observed $\tilde{C}_\ell^{XY}$ as:
\begin{equation}\label{eq:IMxIM_obs}
\tilde{C}_\ell^{\rm IM,IM}(z_i,z_j) = \mathcal{B}(z_i)\mathcal{B}(z_j)C_\ell^{\rm IM,IM}(z_i,z_j)+C_{\ell}^{\mathrm{N,IM}}
\end{equation}
and
\begin{equation}
\tilde{C}_\ell^{\rm IM,GW}(z_i,z_j) = \mathcal{B}(z_i)C_\ell^{\rm IM,GW}(z_i,z_j)
\end{equation}
where the $\mathcal{B}^X(z_i)$ encodes the signal suppression at scales smaller of the FWHM of the beam  $\theta_{B}$. In single-dish configuration $\theta_{B} \sim  1.22 \lambda/D_{d}$, thus implying a more severe suppression of the signal at lower frequencies:
\begin{equation}
    \mathcal{B}(z_i)=
    \exp[-\ell(\ell+1)(\theta_B(z_i)/\sqrt{16\ln2})^2].
\end{equation}
In equation \eqref{eq:IMxIM_obs}, the term $C_{\ell}^{\mathrm{N,IM}}$ indicates noise sources (see also equation \eqref{eq:tilde_Cls}). For the HI case we consider the intrinsic noise of the instrument $C_{\ell}^{\mathrm{instr}}$ and the residual error due to the procedure of cleaning the cosmic IM signal from the bright foreground emission $C_{\ell}^{\mathrm{fg}}$:
\begin{equation}
    C_{\ell}^{\mathrm{N,IM}} = C_{\ell}^{\mathrm{instr}}+C_{\ell}^{\mathrm{fg}},
\end{equation}
whereas the shot-noise is instead a very subdominant component (see e.g., \cite{Castorina+17:HI,Villaescusa+18:ingr}). In the following paragraphs we describe how these noise sources are treated.

\subsubsection*{Instrumental noise}
The experiment setup considered in this work is IM performed in single dish \cite{SKA_redbook, Santos+17} mode and with a collection of $N_d$ dishes. The noise angular power spectrum for this case is given by (see e.g., \cite{Bull15:IM,Santos15:SKA,Santos+17}):
\begin{equation}\label{eq:noise_IM}
C_{\ell}^{\mathrm{instr}}= \sigma_T^2 \theta_B^2. 
\end{equation}
The single-dish rms noise temperature $\sigma_{T}$ writes as
\begin{equation}
\sigma_{T} \approx \frac{T_{\mathrm{sys}}}{\sqrt{n_{\mathrm{pol}} B t_{\mathrm{obs}}}} \frac{\lambda^{2}}{\theta_{\mathrm{B}}^{2} A_{e}} \sqrt{S_{\mathrm{area}} / \theta_{\mathrm{B}}^{2}} \sqrt{\frac{1}{N_{\mathrm{d}}}}.
\end{equation}
Since the beam FWHM of a single dish is $\theta_{B} \sim  1.22 \lambda/D_{d}$, one gets $\lambda^2/A_e \sim \lambda^2/D_d^2 \sim \theta_{B}^2$, where $D_d$ is the diameter of a single dish, $A_e$ is the effective collecting area of the dish and $\lambda= \lambda(z)$ is the observed wavelenght of the redshifted 21cm signal emitted at $z$: $\lambda(z)=\lambda_{21cm}(1+z).$ From this, one can write:
\begin{equation}
C_{\ell}^{\mathrm{instr}}(z_i) \approx \Biggl(\frac{T_{\mathrm{sys}}}{T_b(z_i)\sqrt{n_{\mathrm{pol}} B t_{\mathrm{obs}} N_d}} \sqrt{\dfrac{S_{\mathrm{area}}}{\theta_{\mathrm{B}}^{2}}}\dfrac{1}{T_b(z_i)}\Biggl)^2\theta_{B}^2.
\end{equation}
Following SKAO-MID prescriptions, we have the following parameters values: $T_{\rm sys} = 28K$ for the system temperature, $B=20\cdot 10^6 Hz$ for the bandwidth, $t_0 = 5000 h = 1.8 \cdot 10^{7} s$ for the observation time, $N_d=254$ for the total number of dishes, $S_{\mathrm{area}}=20000 deg^2$ for the total surveyed area, $A_e=140 m^2$ and $D_d=15m$. Notice the normalization to the mean brightness temperature at the center of the redshift bin $T_b(z_i)$, needed to retrieve a dimensionless power spectrum to be added to the theoretical dimensionless one in order to estimate the observed $C_\ell$s, according to equation \eqref{eq:tilde_Cls} . Numerical values are taken from table 2 of reference \cite{Santos15:SKA}. Even though $S_{\mathrm{area}}=20000 \: deg^2 \sim f_{\rm sky}=0.5$ is the official expected value of sky coverage, for the purpose of this work we consider also different values of $f_{\rm sky}$.

Finally, we remark that the $C_{\ell}^{\mathrm{instr}}(z_i)$ noise component is expressed as function of one single redshift because we assume that it is de-correlated among different bins, affecting only auto-correlations.

\subsubsection*{Foregrounds}
The presence of strong foregrounds is one of the central challenges of IM, currently preventing a detection in auto-correlation of the signal (see e.g.,~references \cite{Switzer+2013,Wolz+21}). Such detection should instead be possible for an IM survey with SKAO telescope due to improvements in the signal-to-noise, to a larger scanned sky patch and to the larger frequency band. Nevertheless, the cleaning procedure will not be perfect and the recovery of the pristine HI signal will still be partially complicated by the foreground emission. This effect has been studied with simulations with various degrees of complexity \cite{Alonso14:fogs,Carucci:2020enz,Cunnington:2020njn,Matshawule2020,Soares_2021}. For the purposes of this work, we quantify the residual error that could be expected after a foregrounds removal procedure adding  a noise term $C_\ell^{\mathrm{fg}}$ to the theoretical ${C}_\ell$s, for the $\mathrm{IM \times IM}$ components and for any redshift bins combination. 
Note that, since we focus on the angular power spectrum, we do not model the well known foreground cleaning effect of removing too much power at large scales along the line-of-sight. Our noise term is only a residual systematic accounting for the difficulties in cleaning large spatial scales.
We model the $C_\ell^{\mathrm{fg}}$ term as 
\begin{equation}\label{eq:noise_fog}
C_\ell^{\mathrm{fg}} = K^{\mathrm{fg}}\cdot F(\ell),
\end{equation}
where $K^{\mathrm{fg}}$ is a normalization constant determining the overall amplitude of the residual foregrounds related errors and $F(\ell)$ encodes the scale-dependency.  We write this term as
\begin{equation}\label{eq:F_fid}
F(\ell) = \dfrac{1}{f_{\mathrm{sky}}} A e^{b \ell^c},
\end{equation}
accounting for a larger effect of the cleaning of the signal at larger scales and simply fitting this expression to results of \cite{Alonso14:fogs} (middle-left panel of their figure 3), obtaining $A\sim0.129, \: b\sim-0.081, \: c\sim0.581$. This procedure introduces an error of around 12\% at $\ell\sim 2$ and 4\% at $\ell \sim 100$ (for $f_{\mathrm{sky}}=1.0$). It is also possible to define the variance of this systematic error (see e.g.,~\cite{Camera16:sys}):
\begin{equation}
\sigma_{\mathrm{sys}}^{2}=\int \frac{\mathrm{d} \ln \ell}{2 \pi} \ell(\ell+1)\left|C_\ell^{\mathrm{fg}}\right|.
\end{equation}
Setting the value of the overall normalization factor $K^{\mathrm{fg}}$ to an average value of all the $C_\ell^{\mathrm{IM,IM}}(z_i,z_j)$ components:
\begin{equation}
K^{\mathrm{fg}} = \left\langle C_\ell^{\mathrm{IM,IM}}(z_i,z_j)\right\rangle,
\end{equation}
and considering our redshift binning, we obtain a fiducial value of
\begin{equation}\label{eq:Kfog}
K^{\mathrm{fg}} \simeq 6\cdot 10^{-7}.
\end{equation}  
This leads to a variance of $\sigma_{\mathrm{sys}}^2\simeq 3\cdot 10^{-7}$, in agreement with what assumed in \cite{Camera16:sys}.

\begin{figure}
	\centering
	\includegraphics[width=1.0\linewidth]{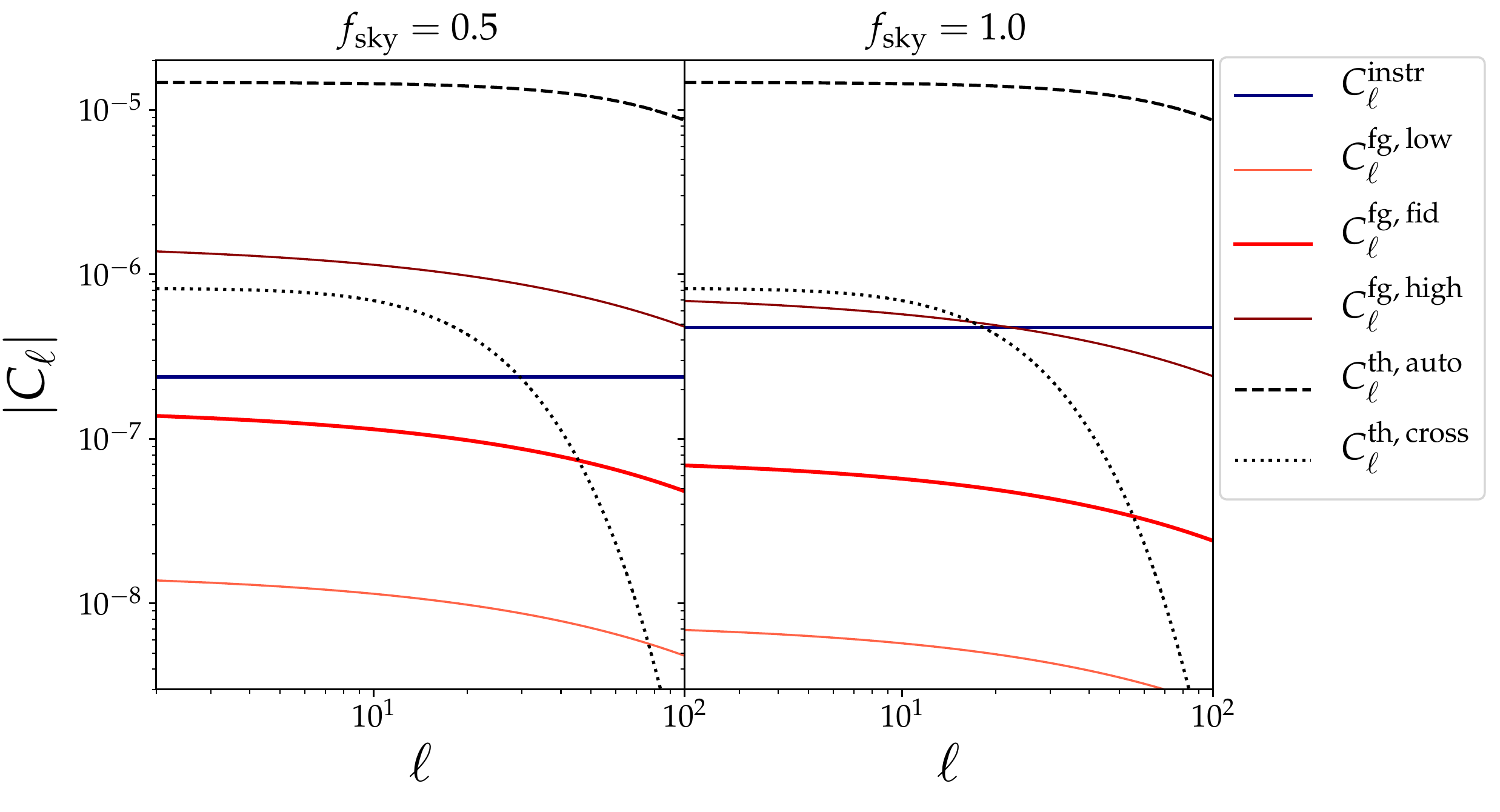}
	\caption{Comparison between HI angular power spectra $C_\ell$s. In black: theoretical power spectra $C_\ell^{\rm th}$ from \texttt{Multi\_CLASS} for the $\mathrm{IM \times IM}$ case, auto-correlating the $20^{\rm th}$ redshift bin ($2.4<z<2.5$) with itself (dashed line) and cross-correlating the $15^{\rm th}$ redshift bin ($1.9<z<2.0$) with the $20^{\rm th}$ (dotted line), as exemplificative case. In blue: instrumental noise power spectrum for SKAO-MID survey at redshift $z_{\mathrm{mean}}=2.5$. In different shades of red: foreground cleaning noise power spectra, with amplitudes equal to $K^{\mathrm{fg}}$ ($C_\ell^{\mathrm{fg, \: fid}}$), $0.1 \cdot K^{\mathrm{fg}}$ ($C_\ell^{\mathrm{fg, \: low}}$) and $10 \cdot K^{\mathrm{fg}}$ ($C_\ell^{\mathrm{fg, \: high}}$). Left(right) panel corresponds to $f_{\mathrm{sky}}=0.5(1.0)$.}
	\label{fig:Cl_err_IM}
\end{figure}

In figure \ref{fig:Cl_err_IM} we compare the resulting $C_\ell^{\mathrm{fg}}$ for different amplitudes $K^{\mathrm{fg}}$ to the contribution from instrumental noise (at $z = 2.5$ by way of example). The contamination term from foreground removal is always dominant and stronger at low multipoles. We report in the figure also an example of the auto angular power spectrum of the HI signal and a cross power spectrum between two different bins. 

We have repeated our analysis using different forms for  \eqref{eq:F_fid} and with different amplitudes, and found a change in the Fisher estimated errors always below $\sim 10-15\%$. Therefore, this would not alter the conclusions reached. Note that our residual foreground contribution is neither frequency dependent nor considers possible coupling between the different scales. These should be secondary effects, especially assuming a full sky survey and a blind cleaning approach to cleaning \cite{Alonso14:fogs,Spinelli2021}.

\section{Gravitational waves statistical redshift distribution}
\label{sec:dNdz_reconstr}
The first application of the $\mathrm{GW\times IM}$ cross-correlation we present in this work is the possibility to obtain a statistical determination of the redshift distribution of GWs detected by laser interferometers. Those detections in fact provide information only on the chrip mass of the system and the (GW) luminosity distance, therefore redshift information is only a derived quantity, based on astrophysical and cosmological assumptions~\cite{Schutz:1986, Bertacca:GWDL}. We rely on the idea that a given tracer whose redshift distribution is well defined can help in calibrating that of a second tracer through studying the cross-correlation of the two. This was already done in several works with techniques such as correlations or the so-called Clustering Based Redshift estimation (see e.g.,~\cite{Newman_2008,Benjamin_2010,Matthews_2010,Schmidt_2013,menard2013clustering,McQuinn13:cbr,Kovetz16:cross,Choi_2016,Scottez_2016,Rahman_2016,Johnson_2016,Alonso17:lssxim,Daalen_2018,Cunnington18:lssxim,Alonso_2021,Mukherjee:gwxlss2,Mukherjee:gwgr}). We mainly follow the methodology of reference \cite{Alonso17:lssxim}, in which the authors perform a redshift calibration of a photometric sample of galaxies through cross-correlation with a spectroscopic sample.

Our case is equivalent, but we make use of the IM of the HI to calibrate GWs events. In fact, GW detected events are characterized by a big uncertainty on their localization due to poor angular resolution, whereas IM provides a refined sliced redshift information about the line under study. Assuming the progenitors of the merging BBHs have astrophysical origin, they are expected to trace very well the underlying distribution of the LSS, which is also well traced by the HI distribution. Combining these two tracers together is expected to improve the redshift localization of the poorly known one (GW) thanks to the refined information coming from other, much better localized (IM). We stress the fact that performing this analysis with resolved photometric galaxy samples instead of IM would not be feasible, due to their lower redshift resolution.

Let us stress again that this methodology aims at calibrating the statistical distribution of a tracer which is poorly localized: this is why we only consider BHBH mergers without taking into account also Neutron Stars (NS) binaries or BHNS systems. Indeed, these latter types of systems can likely be matched to an electromagnetic counterpart, allowing for a well precise localization of the binaries, much more competitively than what the method explored in this section can accomplish.

Given our fiducial redshift distribution for GWs (as described in section \ref{sec:tracer_GW}), we re-model it as a piecewise (PW) function. Each piece has width equal to the IM bin width $\Delta z^{\mathrm{IM}}$, so that the total number of pieces is equal to $N_{\mathrm{bins}}^{\mathrm{IM}}=30$ and each of them perfectly overlaps with a specific IM redshift bin. The overall amplitude of the PW function in the $i^{th}$ bin is indicated as $A_i$. 

Following the formalism described in section \ref{sec:formalism_Fisher}, we perform a Fisher matrix analysis considering the following set of Fisher parameters: the 30 amplitudes $\{A_i\}$ of the GWs PW redshift distribution and \{$\ln 10^{10}A_s, n_s, \bar{b}_{\mathrm{GW}}, \bar{b}_{\rm HI}, K^{\mathrm{fg}}$\}  (for a total of 35 parameters)\footnote{The fiducial parameters values we use in this pipeline are: $\{\ln 10^{10}A_s, n_s, \bar{b}_{\mathrm{GW}}, \bar{b}_{\rm HI}, K^{\mathrm{fg}}\} = \{3.098,0.9619,2.166,1.851,6\cdot 10^{-7} \}$. The cosmology parameters values are taken from Planck \cite{Planck:XIII}, the biases values are obtained applying equation \eqref{eq:bias_def} in the considered redshift range $[0.5,3.5]$ and the $K^{\mathrm{fg}}$ value derives from equation \eqref{eq:Kfog}. Finally, the fiducial values for the amplitudes $\{A_i\}$ are given by the amplitudes of the full GWs redshift distribution at the corresponding redshifts.}. The spectral index and amplitude $n_s$ and $A_s$ are introduced in the pipeline in order to account for a possible cosmology dependence of our results. A fully cosmology based pipeline (including also the dark energy parameters $\{w_0,w_a\}$) will be explored in section \ref{sec:DE}. We set Planck priors on $\{\ln 10^{10}A_s, n_s\}$ \cite{Planck:XIII}.

Note that there is a disadvantage in performing $\mathrm{GW \times IM}$ instead of $\mathrm{LSS \times IM}$ (such as in reference \cite{Alonso17:lssxim}) and constraining the GWs redshift distribution is more difficult than that of photometric galaxy samples. Firstly, the sum over multipoles of equation \ref{eq:Fisher_matrix} stops at $\ell_{max}=100$ (for ET), while galaxy surveys provide a much higher angular resolution (e.g.,~$\ell_{max} \sim 2000$ in \cite{Alonso17:lssxim}). Also, less objects are detected when considering GWs signals from BHBH mergers, inducing a more relevant shot noise contribution, which is translated into higher error bars.

\begin{figure}
	\centering
	\includegraphics[width=1.0\linewidth]{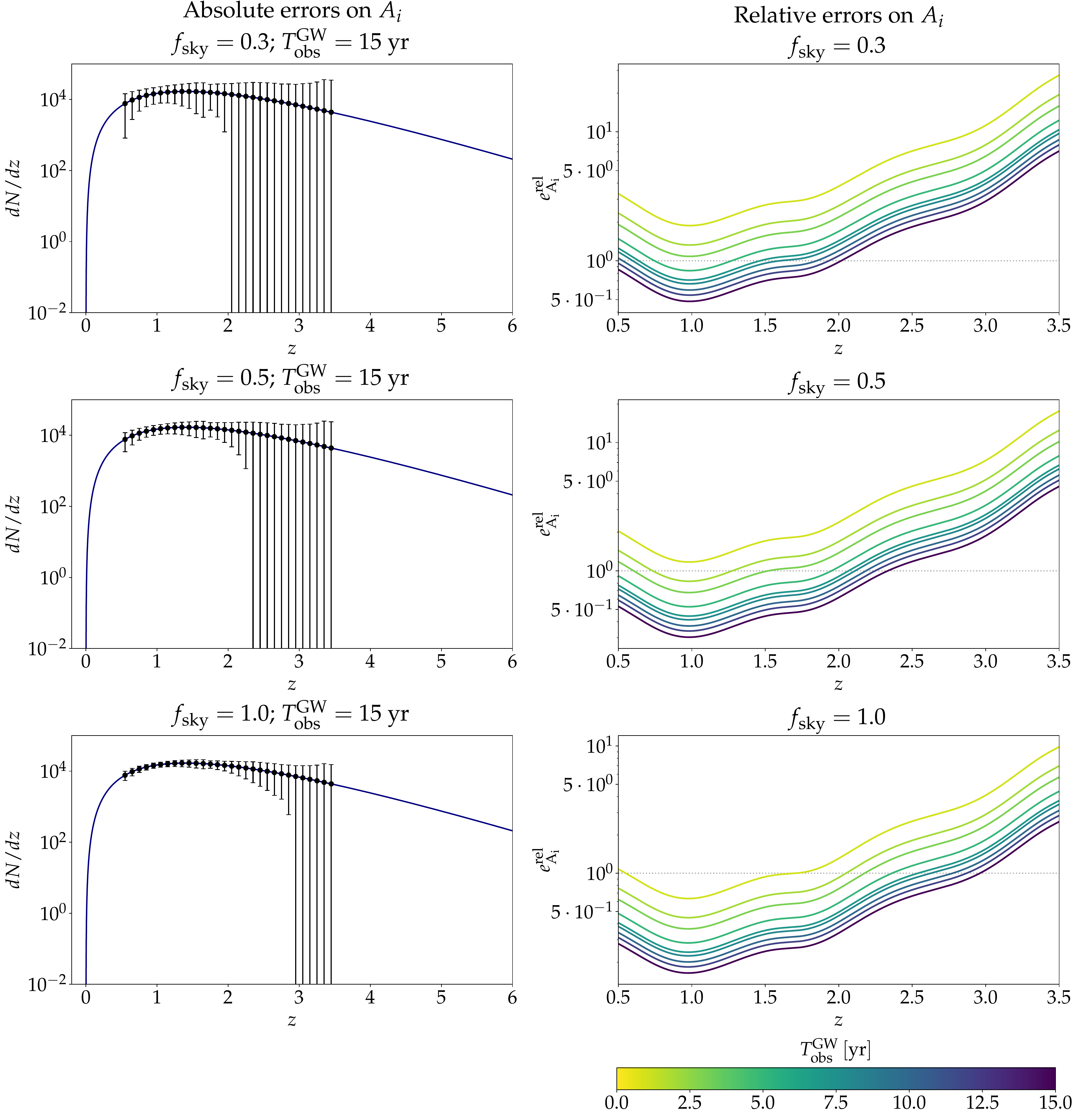}
	\caption{\textit{Left:} GW fiducial redshift distribution (in blue) with error-bars on the amplitude parameters $A_i$. \textit{Right:} Fisher estimated relative errors $e^{\rm rel}_{\rm A_i}$ on the amplitude parameters $A_i$. Different values of $T_{\mathrm{obs}}^{\rm GW}$ (from 1 yr to 15 yr) and $f_{\mathrm{sky}}$ ($f_{\mathrm{sky}}$ = 0.3, 0.5, 1.0) are provided.}
	\label{fig:Ai_errors_full}
\end{figure}

From the Fisher estimated error $\sigma_{\rm A_i}$ on the amplitudes $A_i$ one can directly compute a relative error given by the fraction between the error and the fiducial value $A_i$:
\begin{equation}
e^{\rm rel}_{\rm A_i} = \sigma_{\rm A_i}/A_i.
\end{equation}
In figure \ref{fig:Ai_errors_full} we show the error bars on the 30 GW redshift distribution amplitudes $A_i$ with absolute errors (left panels) and relative errors (right panels) for various combinations of $T_{\mathrm{obs}}^{\rm GW}$ and $f_{\mathrm{sky}}$. We recall that the $f_{\mathrm{sky}}=0.5$ value is the fiducial scenario for SKAO-MID, whereas cases for lower values of $f_{\mathrm{sky}}$ can be considered conservative, and higher values can be thought as a limiting case for future experiments. It is possible to see that for enough high values of $T_{\mathrm{obs}}^{\rm GW}$ and $f_{\mathrm{sky}}$ the error-bars are relatively small and the relative error is below unity. This takes place mainly at lower redshifts, whereas increasing $z$ the error size becomes large and the relative error is raised above unity. Overall, it is possible to see that redshift calibration of GWs events is quite effective in the low-medium redshift range, whereas at higher redshift very high values of $T_{\mathrm{obs}}^{\rm GW}$ and $f_{\mathrm{sky}}$ would be required, when possible. 

We conclude noting one fundamental peculiarity of our approach, which is basically ``cosmology-agnostic'', in the sense that it solely relies on the cross-correlations between GWs (in large z bins, to account for redshift localization uncertainties) and IM (in appropriately small bins). It is independent from the underlying ``true'' cosmology, since we have only assumed fiducial cosmological parameters to perform our Fisher forecasts, but no assumptions in this regards will be needed when working with actual data (thanks to the large width of the considered GWs redshift bins). In addition, note also that in our Fisher analysis we allow for variation of the 30 amplitudes on the GWs redshift distribution: this essentially corresponds to not imposing any prior on the shape of this distribution, taking out any astrophysical assumption that would have imprints on the distribution shape. Forecasts obtained in this way are more pessimistic than other works in current literature (see e.g., \cite{Ng_2021}) and this is due to the fact that a very general approach is being taken here, without any cosmology/astrophysics priors. Finally, let us stress again that the aim of this application is to calibrate the \textit{statistical} distribution of GWs sources, i.e., improving the knowledge on the disposition along redshift of the whole ensemble. We do not aim at better localizing each single event. For this reason, we do not need to take into account the redshift error of each source, as long as the GW redshift binning is large enough the compensate any possible uncertainty in this regard.

\section{Cosmological constraints: dynamical dark energy}\label{sec:DE}
The $\mathrm{GW\times LSS}$ cross-correlation, as a tracer of the matter density field, can provide information on a variety of cosmological parameters. Regarding the $\mathrm{GW \times IM}$ observable, given the specific redshift range and peculiarities of the expected signal, there might be some particular model or parameter that will be optimally tested by it.

In this section we present an example of one of such measurements, which is the possibility of constraining dark energy parameters; in particular, we will focus on parameters describing the redshift evolution of its equation of state.

Starting from the Einstein field equations, which describe how gravity behaves due to the presence of mass-energy (and how this moves given the space-time curvature), one can write
\begin{equation}\label{eq:einstein}
    G_{\mu \nu}= T_{\mu \nu} \, ,
\end{equation}
with the appropriate choice of conventions.
Here $G_{\mu \nu}$ and $T_{\mu \nu}$ are the so called Einstein tensor and stress-energy tensor respectively.
In order to account for the observed accelerated expansion of the Universe, in a General Relativistic framework, equations \eqref{eq:einstein} have to be modified substituting $T_{\mu \nu}$ with $T'_{\mu \nu}=T_{\mu \nu}+T^{\rm DE}_{\mu \nu}$, where $T^{\rm DE}_{\mu \nu} = -\Lambda g_{\mu \nu} $
(with $g_{\mu \nu}$ the metric tensor) and $\mathrm{DE}$ stays for dark energy. This can be in the form of a cosmological constant or an additional field.
Among the many possible proposed models and deviations from the cosmological constant (originally inspired by vacuum energy), a possibility is to investigate deviations from its equation of state (eos) being constant with redshift.
The eos is defined as $w = \dfrac{p}{\rho}$ (where $p$ and $\rho$ are respectively the pressure and energy densities of the fluid). If the DE behaviour is described by a cosmological constant, then $w=-1$, whereas the relation
\begin{equation}
    w(a) = w_0 + w_a(1-a)
\end{equation}
generally describes the case of time evolution. 
Current measurements from CMB and galaxy clustering are consistent with a cosmological constant~\cite{planck:2018, boss:de}, but many models predict small departures from the cc case (see e.g.,~\cite{Giudice:2021, Heisenberg:swampland}), therefore this remains an important part in the efforts toward a better understanding of the cosmological model.

Here we investigate how well the DE parameters $\{w_0,w_a\}$ can be constrained through $\mathrm{GW \times IM}$ cross-correlations. We include in our Fisher analysis pipeline summarized in section \ref{sec:formalism_Fisher} the following parameters: $\{w_0, w_a, \omega_{\rm cdm}, \omega_{\rm b}, 100\theta_s, \ln 10^{10}A_s, n_s, \bar{b}_{\mathrm{GW}}, \bar{b}_{\rm HI}, K^{\mathrm{fg}}\}$ (for a total of 10 parameters)\footnote{The fiducial parameters values we use in this pipeline are: $\{w_0, w_a, \omega_{\rm cdm}, \omega_{\rm b}, 100\theta_s,\ln 10^{10}A_s, n_s, \bar{b}_{\mathrm{GW}}, \bar{b}_{\rm HI},\\ K^{\mathrm{fg}}\} = \{-1.0,0.0,0.12038,0.022032,1.042143,3.098,0.9619,2.166,1.851,6\cdot 10^{-7} \}$. As in the previous section, the cosmology parameters values are taken from Planck \cite{Planck:XIII}, the biases values are obtained applying equation \eqref{eq:bias_def} in the considered redshift range $[0.5,3.5]$ and the $K^{\mathrm{fg}}$ value derives from equation \eqref{eq:Kfog}.}. The differences with respect to the analysis of section \ref{sec:dNdz_reconstr} consist in the usage of the full GWs redshift distribution (in place of its piece-wise approximation) and the addition of other cosmology related Fisher parameters: $w_0$, $w_a$, the cold dark matter physical density $\omega_{\rm cdm}=\Omega_{\rm cdm}h^2$, the baryon physical density $\omega_{\rm b}=\Omega_{\rm b}h^2$ and the angular scale of the sound horizon at decoupling $100\theta_s$. We set a Planck prior on the $\{\omega_{\rm cdm}, \omega_{\rm b}, 100\theta_s, \ln 10^{10}A_s, n_s\}$ parameters, unless where stated otherwise.

\begin{figure}
	\centering
	\includegraphics[width=1.0\linewidth]{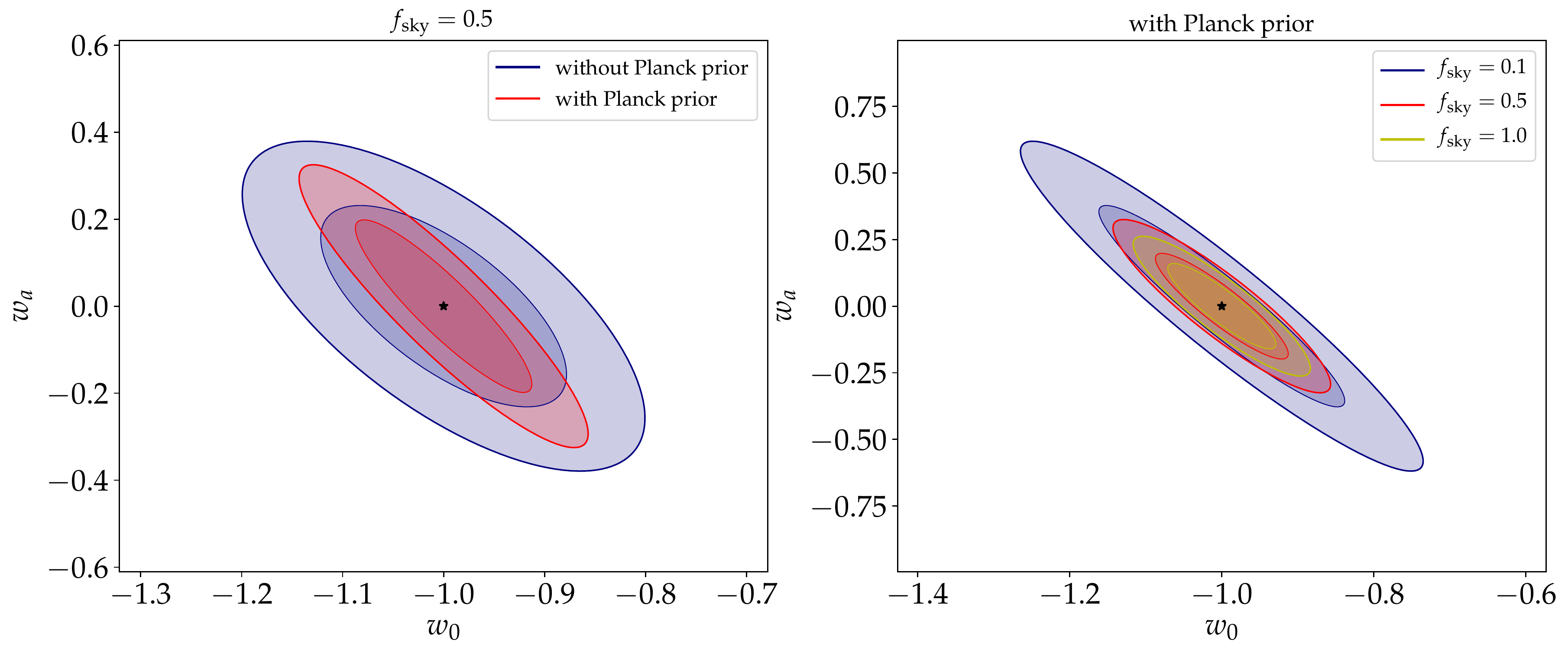}
	\caption{Contour plots for the DE parameters $\{w_0,w_a\}$ at $1\sigma$ and $2\sigma$ confidence levels, for fixed $T_{\rm obs}^{\rm GW}=10{\rm yr}$. \textit{Left:} forecasts for fixed $f_{\rm sky}=0.5$. With and without applying a Planck prior on other cosmology parameters. \textit{Right:} forecasts for different values of $f_{\rm sky}$ (as legend). A Planck prior on all other cosmology parameters is applied.}
	\label{fig:DE_contours}
\end{figure}

In figure \ref{fig:DE_contours} we provide forecasts for the constraining power on the dark energy parameters $\{w_0,w_a\}$ for the experiments considered in this work, including or not Planck priors and for different values of $f_{\rm sky}$. On the right panel we can see the improvement that would come by increasing the fraction of the sky surveyed. Overall, we can notice that the predicted constraints are qualitatively in agreement with other studies, such as e.g.,~reference \cite{Raccanelli16:radio} in which a study of the cross-correlation between GWs and radio galaxy surveys is performed or reference \cite{Bull15:IM} in which forecasts are obtained for HI intensity mapping experiments. 

We can also see that constraints are approximately comparable to the BOSS+Planck results available in the literature (see e.g.,~\cite{boss:de}) for what concerns the errors on the $\{w_0,w_a\}$ parameters, with $w_0$ constraints slightly weaker. Comparing our results with those from the Euclid collaboration, we can see that (see e.g., \cite{blanchard2020euclid}) our constraints are approximately comparable or slightly less competitive when considering combinations of all probes tested by Euclid (weak lensing and spectroscopic/photometric galaxy clustering), but more stringent with respect to Euclid forecasts obtained through one single probe. A similar comparison can be done when looking at forecasts for the Vera Rubin Observatory (LSST, see e.g., \cite{LSST_DE}): our forecasts are more optimistic than those from the LSST (up to $50\%$) when considering just the single isolated probes testable by the survey, but become less competitive when their combination in the LSST is exploited. All in all, although the forecasts we obtain with this methodology are not competitive with the maximum potential that other surveys (such as those mentioned above) can achieve, they can still provide an alternative way to test these scenarios. Cross-correlations are well know for helping in reducing systematics (see e.g.,~\cite{Bonaldi:sys}). For this reason, these measurements will provide a very useful cross-check to available results, as they will come from cross-correlating two very different datasets and will be affected by much less and different systematics.

\section{Astrophysical vs. primordial origin of merging black hole binaries}
\label{sec:pbhvsastro}
In this section we tackle the issue of using $\mathrm{GW \times IM}$ for determining the origin of the progenitors of merging BBHs, distinguishing between the possibility that they originate from the end-point of stellar evolution, or that they are Primordial Black Holes (PBHs) generated in the early universe.

PBHs were first theorized a few decades ago \cite{hawking:pbh, carr:pbh} when it was proposed that some over-dense regions in the primordial universe could reach the threshold for gravitational collapse and, in some regions, form black holes.
Several formation mechanism have been proposed in literature, such as the collapse of cosmic string loops or domain walls \cite{Polnarev88, HAWKING1989237,WICHOSKI1998191,BEREZIN198391,Ipser84}, the collapse of large fluctuations at inflation \cite{ivanov:pbhfrominflationI, Bellido:pbh, ivanov:pbhfrominflationII}, bubble collisions \cite{Crawford82,LA1989375}, but the mainstream hypothesis is that they originate from large perturbations in the primordial curvature power spectrum (that went outside of the horizon during inflation) right after horizon re-entry; there has been an intense activity in the community in the last few years on the relation between the primordial power spectrum and PBHs (see e.g.,~\cite{musco:2005, cole:2018, kalaja:2019, musco:2019, young:2019, byrnes:2019, satopolito:2019, Munoz:2017,Motohashi:2017,gow:2020,musco:2021, byrnes:2021,inomata2021primordial}).

The interest towards this type of compact objects revived after the first detection of GWs from the merger of two massive black holes  \cite{abbott:firstligodetection,abbott:firstligodetectionproperties}, when it was proposed that their progenitors might have primordial origin and even constitute a non negligible fraction of the dark matter, reviving the ``PBHs as dark matter'' hypothesis (e.g.,~\cite{bird:pbhasdarkmatter,clesse:pbhmerging}).
There is still no conclusive agreement on the possibility that stellar-mass PBHs exist in sufficient abundance to make up for a considerable part of the dark matter (see e.g.,~\cite{brandt:ufdgconstraint, Munoz:2016, green:2016, zumalacarregui:supernovaconstraint, murgia:pbh, AliHaimoud:PBHmergerrate, deluca:nanograv, mukherjee:abhpbh} for some studies and constraints in this mass range), but confirmation (or exclusion) of their sheer existence would represent a big step in our understanding of the Universe. In fact, detecting even one PBH would provide invaluable information on the physics of the early Universe on scales otherwise inaccessible to standard cosmological measurements; moreover, it was recently shown that PBHs and WIMP DM are incompatible~\cite{adamek:wimp_pbh}, so that the observation of one PBH would rule out the main DM candidate model.

Therefore, it will be extremely important to add another probe for the possible detection of the presence of PBHs and to understand their formation channels and merger rates. Measurements of $\mathrm{GW \times IM}$ will then add further information on this issue, on redshift ranges and with data sets complementary to what is and will be available otherwise.
We refer the reader to PBH reviews such as e.g.,~\cite{Carr:2016_rev, Sasaki_2018, Green:review, Carr_Silk_2018,Carr20:review} for more details.

The idea on top of which we build for our study follows the same logic of references \cite{raccanelli:pbhprogenitors, Scelfo18:gwxlss}: approaching the problem in a statistical way, we know that GWs from merging BBHs trace the underlying matter distribution in ways that depend on their origin (see later for more details) and consequently would correlate with the LSS - and the HI distribution, which is a tracer of the LSS - in different ways.

The relation between observables and the underlying matter distribution is encapsulated in the bias parameter introduced in section \ref{sec:formalism_Cls}.
Crucially, it has been shown~\cite{bird:pbhasdarkmatter, AliHaimoud:PBHmergerrate} that PBH mergers trace halos and the stellar distributions in ways that are different from endpoint of stellar evolution BHs, and in different ways depending on the PBH binary formation mechanism.
This will then assign a different preferred bias $\bar{b}_{\mathrm{GW}}$ for the GWs, which will be the discriminant we can use for our study.
The main features of the scenarios we aim to distinguish through $\mathrm{GW \times IM}$ are sketched in the following section.

\subsection{Progenitors}
\label{sec:scenarios}
In this section we briefly characterize the scenarios (astrophysical and different primordial ones) that we compare. To do so, we introduce the $\Gamma_{\rm pbh}$ parameter, which indicates the fraction of detected merging BBHs with primordial origin (over the total number of observed BBH mergers). 
Assuming the detection of $\mathcal{N}_{\rm tot}$ mergers, of which $\mathcal{N}_{\rm astro}$ have astrophysical origin and $\mathcal{N}_{\rm pbh}$ have primordial origin ($\mathcal{N}_{\rm tot}=\mathcal{N}_{\rm astro}+\mathcal{N}_{\rm pbh}$), the $\Gamma_{\rm pbh}$ parameter is defined as:
\begin{equation}\label{eq:gamma_pbh}
    \Gamma_{\rm pbh} = \mathcal{N}_{\rm pbh}/\mathcal{N}_{\rm tot}
\end{equation}
and spans from $\Gamma_{\rm pbh}=0$ (i.e.,~only astrophysical BBH mergers are detected) to $\Gamma_{\rm pbh}=1$ (i.e.,~only primordial BBHs mergers are detected).

\subsubsection{Astrophysical scenario}
In this case, the progenitors of merging BBHs are formed at the end-point of stellar evolution. All the features of the detected GW events originated from their mergers are already discussed in section \ref{sec:tracer_GW}. GWs from mergers of astrophysical BHs will then highly correlate with large, luminous halos that contain the majority of stars, and, consequently, they would highly correlate with the HI IM signal. The average bias in the redshift range considered here (calculated combining equations \eqref{eq:bias_GWs} and \eqref{eq:bias_def}) is
\begin{equation}\label{eq:bias_astro}
\bar{b}_{\mathrm{GW}}^{\mathrm{ASTRO}} \sim 2.17.
\end{equation}
Finally, by definition this scenario is characterized by a value of $\Gamma_{\mathrm{pbh}}=0$.

\subsubsection{Primordial scenario: ``early'' binaries}
We start our analysis of the $\mathrm{GW \times IM}$ that will be measured if $\Gamma_{\rm pbh} \neq 0$ with the scenario in which the vast majority of primordial black holes binary formation took place in the early universe (see e.g.,~\cite{AliHaimoud:PBHmergerrate,Raidal19:pbh}), while late time formed PBH binaries are assumed to give a negligible contribution. 

In analogy to the correlation of GW with galaxies, the HI distribution is expected to correlate with PBHs mergers from early binaries with a different bias. This is due to the fact that these binaries would form in correspondence of the DM distribution, tracing very well the underlying matter distribution, instead of just tracing locations with massive and luminous halos.
Assuming $\Gamma_{\mathrm{pbh}}=1.0$, GWs should have an associated bias of
\begin{equation}\label{eq:bias_early}
\bar{b}_{\mathrm{GW}}^{\mathrm{PBH}} \sim 1.0
\end{equation}
since they would trace very effectively the underlying matter distribution. We provide in Figure \ref{fig:tracers_pbh} the specifics characterizing GWs events produced in this scenario. See appendix \ref{app:pbh_fid} for further explanations.

\begin{figure}
	\centering
	\includegraphics[width=1.05\linewidth]{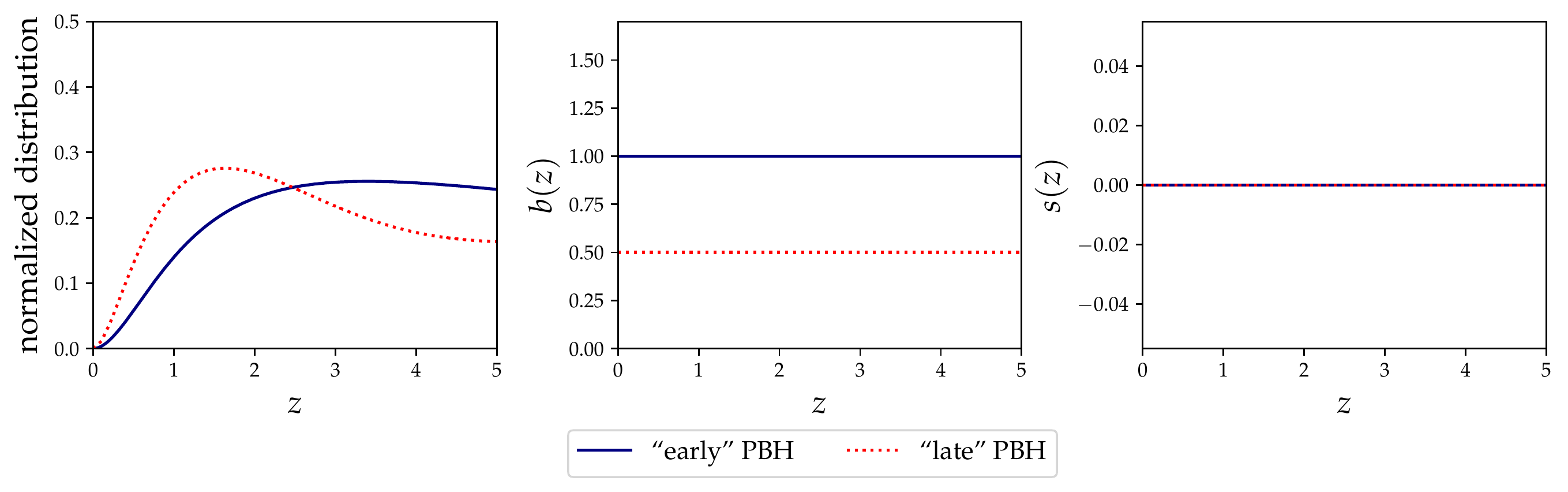}
	\caption{Specifics for the GW tracer in the ``early'' and ``late'' primordial scenarios. \textit{Left:} normalized redshfit distribution. \textit{Center:} bias $b(z)$. \textit{Right:} magnification bias $s(z)$. Magnification bias assumed to be zero for both scenarios (see appendix \ref{app:pbh_fid} for further discussions).}
	\label{fig:tracers_pbh}
\end{figure}

\subsubsection{Primordial scenario: ``late'' binaries}
Under this scenario we again assume that the progenitors of the merging BBHs have primordial origin, but the formation of the binary system itself takes place in the late Universe through a gravitational bremsstrahlung process~\cite{bird:pbhasdarkmatter}.

We assume binary formation happens when two PBHs have a close encounter and their relative velocities are low enough that capture can take place and allows the binary formation.
Given that the velocity dispersion is on average lower within small mass halos, which can not form large quantities of stars and are characterized by low values of the bias parameter, GWs in this case would be anti-correlated with luminous galaxies and (if $\Gamma_{\mathrm{pbh}}=1.0$) GWs would be expected to poorly trace luminous, highly star-forming massive halos and would be characterized by a bias value of
\begin{equation}\label{eq:bias_late}
\bar{b}_{\mathrm{GW}}^{\mathrm{PBH}} \sim 0.5,
\end{equation}
which is typical of the dark, low-mass halos in which this PBHs late binary formation effect would take place \cite{bird:pbhasdarkmatter}.

In Figure \ref{fig:tracers_pbh} we show the specifics characterizing GWs events under this scenario. See appendix \ref{app:pbh_fid} for further explanations.

\subsubsection{Primordial scenario: ``mixed'' binaries}
Finally, we consider a case in which we assume that the progenitors of the merging BBHs have primordial origin, with PBH binary formation taking place through both channels described by the ``early'' and ``late'' scenarios.

The bias parameter value of this model is given by the weighted average of the two values of equations \eqref{eq:bias_early} and \eqref{eq:bias_late}. In particular, as an example case, we assume that around $70 \%$ of the PBHs mergers come from early binaries, whereas around the $30 \%$ is given by mergers of late binaries. This rough estimate comes from taking the lower and upper bounds of the LIGO/Virgo local merger rate estimates and associating them to the ``late'' and ``early'' scenarios respectively.
In fact, even though according to part of the current literature the ``early'' scenario might be the dominant one, the issue is not settled yet. In this case we aim at taking into account both PBHs binary formation channels, considering a non-negligible contribution to the total PBHs merger rate from either mechanism. Assuming $\Gamma_{\mathrm{pbh}}=1.0$, this scenario is then characterized by:
\begin{equation}\label{eq:bias_mixed}
\bar{b}_{\mathrm{GW}}^{\mathrm{PBH}} \sim 0.85.
\end{equation}

\subsection{Forecasts}
We calculate the Signal-to-Noise ratio $S/N$ to quantify how well a fiducial model (astrophysical or primordial) can be distinguished from an alternative one, by looking at the $\bar{b}_{\mathrm{GW}}$ value predicted by the two models:
\begin{equation}\label{eq:SN_astro_vs_pbh}
\left(\frac{S}{N}\right)^2=\frac{\left(\bar{b}_{\mathrm{GW}}^{\text {Alternative }}-\bar{b}_{\mathrm{GW}}^{\mathrm{Fiducial}}\right)^2}{\sigma^2_{\bar{b}_{\mathrm{GW}}^{\mathrm{Fiducial}}}} \, ,
\end{equation}
where $\sigma_{\bar{b}_{\mathrm{GW}}^{\mathrm{Fiducial}}}$ is the Fisher estimated error on $\bar{b}_{\mathrm{GW}}$ in the fiducial scenario. The biases values $\bar{b}_{\mathrm{GW}}^{\mathrm{Alternative}}$ and $\bar{b}_{\mathrm{GW}}^{\mathrm{Fiducial}}$ are those characterizing the models presented in section \ref{sec:scenarios}, depending on which of them is assumed as alternative or fiducial. We obtain $\sigma_{\bar{b}_{\mathrm{GW}}^{\mathrm{Fiducial}}}$ by making use of the same Fisher pipeline (parameters and fiducial values) of the analysis of section \ref{sec:DE}: $\{K^{\mathrm{fg}}, ln10^{10}A_s, n_s, \omega_{\rm cdm}, \omega_{\rm b}, 100\theta_s, w_0, w_a, \bar{b}_{\mathrm{GW}}, \bar{b}_{\rm HI}\}$ for a total of 10 parameters. We set Planck priors on $\{ln10^{10}A_s, n_s, \omega_{\rm cdm}, \omega_{\rm b}, 100\theta_s\}$ \cite{Planck:XIII}. 

We provide forecasts assuming the astrophysical model as fiducial, characterized by $\Gamma_{\mathrm{pbh}}^{\mathrm{FID}}=0.0$. Regarding the alternative models to compare with, we consider a series of mixed astrophysical-primordial scenarios with $\Gamma_{\mathrm{pbh}}^{\mathrm{ALT}}\in [0.0,1.0]$, with a bias given by
\begin{equation}\label{eq:bias_alt}
\bar{b}_{\mathrm{GW}}^{\mathrm{ALT}}=\bar{b}_{\mathrm{GW}}^{\mathrm{ASTRO}}(1-\Gamma_{\mathrm{pbh}}^{\mathrm{ALT}}) + \bar{b}_{\mathrm{GW}}^{\mathrm{PBH}}\Gamma_{\mathrm{pbh}}^{\mathrm{ALT}} \; .
\end{equation}

It is worth noting that this kind of approach may lead to possible degeneracies for some mixed scenarios, i.e., different scenarios combinations might yield the same $\bar{b}_{\rm GW}$. In this eventuality, comparing the bias in specific redshift sub-samples would be enough to break the degeneracy, given that its predicted redshift dependence is different among distinct cases.

\begin{figure}
	\centering
	\includegraphics[width=0.85\linewidth]{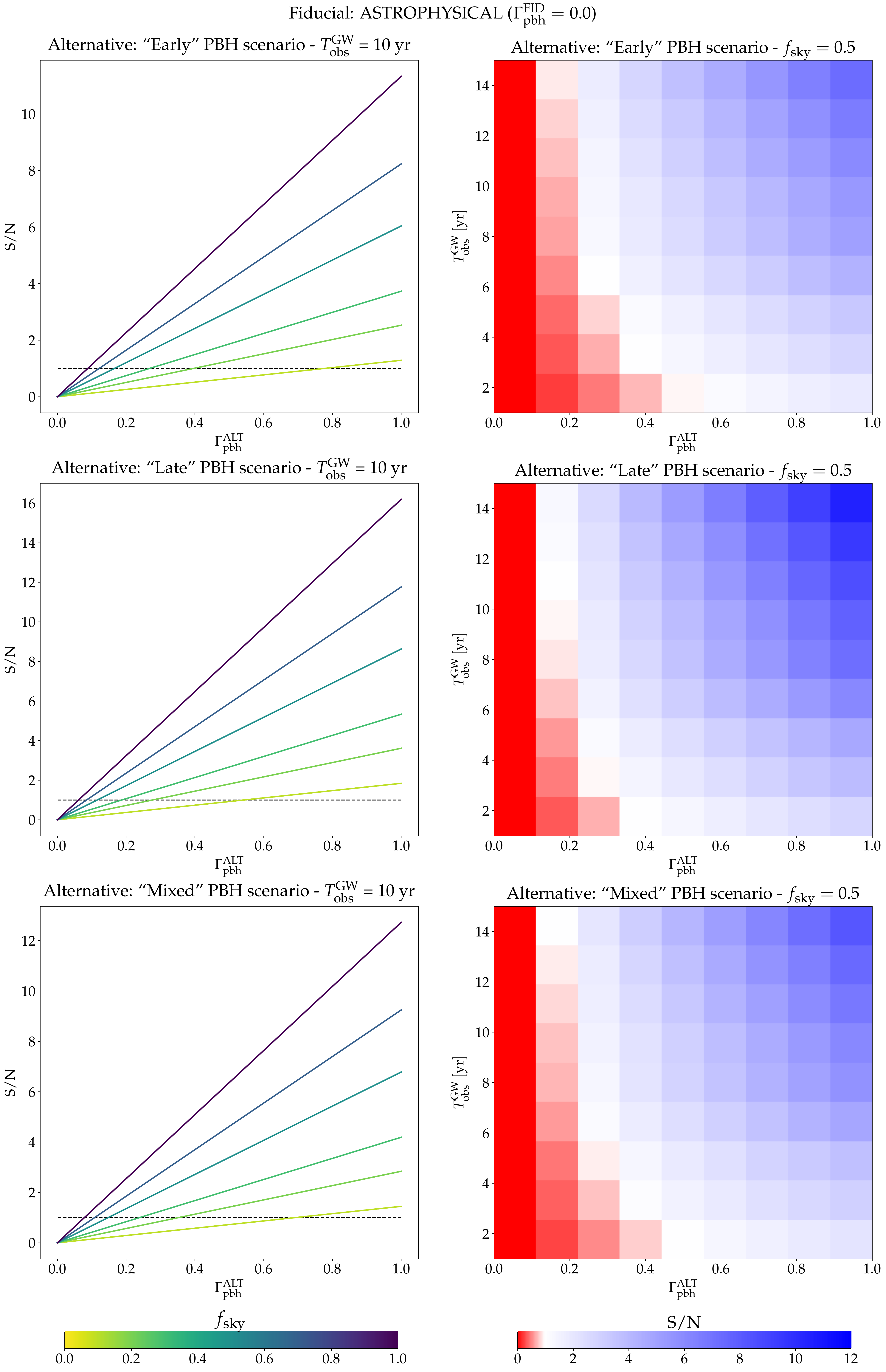}
	\caption{SNR for different values of $T_{\mathrm{obs}}^{\rm GW}$ (from 1 yr to 15 yr), $f_{\mathrm{sky}}$ (from 0.1 to 1.0) and $\Gamma_{\mathrm{pbh}}^{\mathrm{ALT}}$ (from 0.0 to 1.0), assuming the astrophysical model as fiducial. The alternative model assumed is the ``Early'', ``Late'' and ``Mixed'' primordial scenario in the top, center and bottom panel, respectively. The colorbar of the right-side plots is normalized to white at $S/N=1$.}
	\label{fig:SN_astro_fid}
\end{figure}

We provide in figure \ref{fig:SN_astro_fid} SNR estimates from equation \eqref{eq:SN_astro_vs_pbh} for a series of values of $f_{\mathrm{sky}}$ and $T_{\rm obs}^{\rm GW}$ assuming the astrophysical scenario as fiducial and comparing it with the three different primordial scenarios (``early'', ``late'' and ``mixed'') described in section \ref{sec:scenarios}.
On the left panel we show the SNR as a function of $\Gamma_{\mathrm{pbh}}^{\mathrm{ALT}}$, where the color code indicates the fraction of the sky observed. On the right panels we present the SNR obtainable (color coded) as a function of both the observation time and the fraction of primordial BBHs in the alternative model.
In both columns, from top to bottom results for the three scenarios presented above are provided.

It can be seen that for large enough fractions of the sky and observation times, 
results are very promising, providing a $S/N$ well above unity.
In fact, it would be possible to distinguish, at a few sigma, a purely astrophysical model from an alternative model made up of similar relative abundances between astrophysical and primordial BBHs ($\Gamma_{\mathrm{pbh}}^{\mathrm{ALT}} \sim 0.5$) with a few years of observations and $f_{\mathrm{sky}}\sim 0.5$.
In addition, an alternative model with low values of $\Gamma_{\mathrm{pbh}}^{\mathrm{ALT}}$ (i.e.,~mostly made of astrophysical BHs) such as $\Gamma_{\mathrm{pbh}}^{\mathrm{ALT}} \sim 0.2$, could be detected within 10 years of observation for $f_{\mathrm{sky}}=0.5$, that we stress once again being the fiducial value for the considered SKAO-MID survey. We have also tested mixed scenarios with different relative abundances of early/late-type PBHs to astrophysical BHs, finding as expected no extremely different qualitative behaviours when changing these quantities.

Finally, analogous conclusions can be reached when the assumed fiducial model is a primordial scenario: for completeness, we provide forecasts for the example case of ``early'' primordial scenario assumed as fiducial in appendix \ref{app:pbh_fid}.

\section{Conclusions}\label{sec:conclusions}
In this work, we investigated the cross-correlation signal between gravitational wave catalogs and HI intensity maps that could be potentially measured with future experiments such as the Einstein Telescope and the SKA Observatory.
We extended the range of applications of the publicly available code \texttt{Multi\_CLASS}, by including signal from the IM unresolved HI sources and compute their angular power spectra $C_\ell$s, including all projection effects over a variety of redshift ranges.

We presented three cosmological and astrophysical applications we believe will be particularly exciting for the future.

First, we investigated how well the $\mathrm{GW \times IM}$ cross-correlation can provide model independent and agnostic information on the redshift distribution of resolved binary black hole mergers. Our results show that we will be able to obtain good precision in the inferred statistical redshift distribution of BBH merger number counts. With the experiments considered, we will be able to obtain a precision of order a few tens of \% for redshifts up to $z\approx 1.5$, potentially providing ways to discriminate between different astrophysical models of the binary formation, evolution and mergers.

We stress that our methodology does not make use of any astrophysical model, including BH population, mass function, etc, and is free from assumptions on the values of the cosmological parameters. It will be of particular interest to compare results from our methodology with other approaches that either assume the knowledge of a cosmological model, set priors on the BH population distributions or exploit machine learning techniques (see e.g.,~\cite{Mukherjee:gwxlss2, Ng_2021, canas2021gaus}). The combinations of these methods could lead to an improvement of constraints or, if some inconsistencies will emerge, could provide hints of inaccuracies in standard assumptions.

Measurements of power spectra of tracers of the matter distribution over cosmic epochs will naturally provide also cosmological information; as an example showcasing the potential of the $\mathrm{GW \times IM}$ observable, we forecast constraints on dynamical dark energy parameters. Our results show that the $\mathrm{GW \times IM}$ cross-correlations will provide a fundamental validation of IM-only results, as this latter could be a measurement affected by unknown systematic errors.

As a final application for $\mathrm{GW \times IM}$, we studied how we can use this correlation to detect a primordial component in the BBHs detected through GW from their mergers. We follow the approach of~\cite{raccanelli:pbhprogenitors, Scelfo18:gwxlss} and extend it to HI IM maps, at the same time updating merger rate estimates and binary formation channels with the latest results available in literature. Compared to the case of the correlation with star forming galaxies, using IM allows us to reach higher redshifts and have a very fine tomographic binning. Our results show that with $\mathrm{GW \times IM}$ from the ET and SKAO we will be able to detect the presence of a PBH component down to about 30\% of detected mergers at high Signal-to-Noise ratios.

To conclude, we presented the first study of the cross-correlation between gravitational waves from resolved binary black hole mergers and the HI intensity mapping signal, and investigated some possible interesting applications with predictions from the future Einstein Telescope and SKAO experiments. We believe that this first investigation can open up a plethora of new measurements and possibilities for the scientific community. Moreover, it will be of particular interest to combine our suggested approach with other ones such as additional GWxLSS correlations.
This paper can be seen as part of the ongoing effort to develop multi-tracer approaches, which has an enormous potential in both cosmology and astrophysics, as it allows to test models in ways that would not be possible by looking at single tracers separately.

\section*{Acknowledgments}
We are thankful to Jos\'e Luis Bernal, Kaze Wong and Gabriela Sato-Polito for their helpful advices. We thank Nicola Bellomo, Alessandro Bressan, Phil Bull, Stefano Camera, Jose Fonseca, Ely Kovetz, Suvodip Mukherjee, Julian Mu{\~{n}}oz, Gabriele Parimbelli, Mario Spera for useful discussions. 
GS, MS and MV are supported by the INFN PD51 INDARK grant.
MS acknowledges funding from the INAF PRIN-SKA 2017 project 1.05.01.88.04 (FORECaST).
AR acknowledges funding from Italian Ministry of University and Research (MIUR) through the ``Dipartimenti di eccellenza'' project ``Science of the Universe''.
AL acknowledges funding from PRIN MIUR
2017 prot. 20173ML3WW, `Opening the ALMA window on the cosmic evolution of gas, stars and supermassive black holes', and from the EU H2020-MSCA-ITN-2019 Project 860744 `BiD4BESt: Big Data applications for black hole Evolution STudies'.

\vspace{1cm}
\appendix
\section{Relativistic number counts}
\label{app:deltas}
In this appendix we provide the complete expressions for the relativistic number counts effects introduced in equation \eqref{eq:numbercount_fluctuation}:
\begin{equation}
\begin{aligned}
\Delta_\ell^\mathrm{den}(k,z) &= b_X \delta(k,\tau_z) j_\ell,	\\
\Delta_\ell^\mathrm{vel}(k,z) &=  \frac{k}{\mathcal{H}}j''_\ell  V(k,\tau_z) + \left[(f^\mathrm{evo}_X-3)\frac{\mathcal{H}}{k}j_\ell + \left(\frac{\mathcal{H}'}{\mathcal{H}^2}+\frac{2-5s_X}{r(z)\mathcal{H}}+5s_X-f^\mathrm{evo}_X\right)j'_\ell \right]  V(k,\tau_z),	\\
\Delta_\ell^\mathrm{len}(k,z) &= \ell(\ell+1) \frac{2-5s_X}{2} \int_0^{r(z)} dr \frac{r(z)-r}{r(z) r} \left[\Phi(k,\tau_z)+\Psi(k,\tau_z)\right] j_\ell(kr),	\\
\Delta_\ell^\mathrm{gr}(k,z)  &= \left[\left(\frac{\mathcal{H}'}{\mathcal{H}^2}+\frac{2-5s_X}{r(z)\mathcal{H}}+5s_X-f^\mathrm{evo}_X+1\right)\Psi(k,\tau_z) + \left(-2+5s_X\right) \Phi(k,\tau_z) + \mathcal{H}^{-1}\Phi'(k,\tau_z)\right] j_\ell + \\
&+ \int_0^{r(z)} dr \frac{2-5s_X}{r(z)} \left[\Phi(k,\tau)+\Psi(k,\tau)\right]j_\ell(kr) \\
&+ \int_0^{r(z)} dr \left(\frac{\mathcal{H}'}{\mathcal{H}^2}+\frac{2-5s_X}{r(z)\mathcal{H}}+5s_X-f^\mathrm{evo}_X\right)_{r(z)} \left[\Phi'(k,\tau)+\Psi'(k,\tau)\right] j_\ell(kr).
\end{aligned}
\end{equation}
The quantities introduced above have the following physical meaning: $b_X$ is the bias parameter, $s_X$ is the magnification bias parameter, $f^\mathrm{evo}_X$ is the evolution bias parameter, $r$ is the conformal distance on the light cone, $\tau=\tau_0-r$ is the conformal time, $\tau_z=\tau_0-r(z)$, $j_\ell$, $j'_\ell=\frac{dj_\ell}{dy}$, $j''_\ell=\frac{d^2j_\ell}{dy^2}$ are the Bessel functions and their derivatives (evaluated at $y=kr(z)$ when not explicitly stated), $\mathcal{H}$ is the conformal Hubble parameter, the prime symbol $'$ indicates derivatives with respect to conformal time, $\delta$ is the density contrast in the comoving gauge, $V$ is the peculiar velocity, $\Phi$ and $\Psi$ are Bardeen potentials (see e.g.,~reference \cite{didio:classgal} and references therein).

\section{Astrophysical vs. primordial BBHs: ``early'' primordial scenario as fiducial}\label{app:pbh_fid}
For completeness, in this appendix we show that the optimistic results of section \ref{sec:pbhvsastro}, regarding the possibility of determining the progenitors of merging BBHs, are obtainable also when considering a primordial scenario as fiducial model, aiming at distinguishing it from an alternative astrophysical case.

We consider here the ``early'' primordial scenario and assume the redshift evolution of the merger rate described in reference \cite{Raidal19:pbh}, which is an extension to the model of reference \cite{AliHaimoud:PBHmergerrate}. As done for the astrophysical merger rate, we re-normalize it to the value of $30 \: {\rm Gpc^{-3} yr^{-1}}$, in agreement with local LIGO/Virgo observations. In fact, suppression effects to the merger rate could in principle shift it to agree with experimental values \cite{Raidal19:pbh}. We stress that due to the big uncertainties of both the PBHs merger rate and suppression effects, the state of the art in this field is still in full development. 
	
Given the big uncertainties in the PBH modeling, we assume here a magnification bias with a value of $s_{\rm GW}^{\rm PBH}=0.0$, since almost all sources in the $\mathcal{O}(10 M_\odot)$ mass range would be detected by the ET in our redshift interval (see e.g.,~figure 3 of \cite{Scelfo18:gwxlss}) and a slightly more precise determination of this quantity would be possible only by fixing quantities accompanied by huge uncertainties (mass distribution, suppression model, etc.). We provide in figure \ref{fig:tracers_pbh} the specifics assumed for this scenario, comparing them for completeness with those of the ``late'' primordial case (redshift distribution following prescriptions of reference \cite{bird:pbhasdarkmatter}).

\begin{figure}
	\centering
	\includegraphics[width=0.935\linewidth]{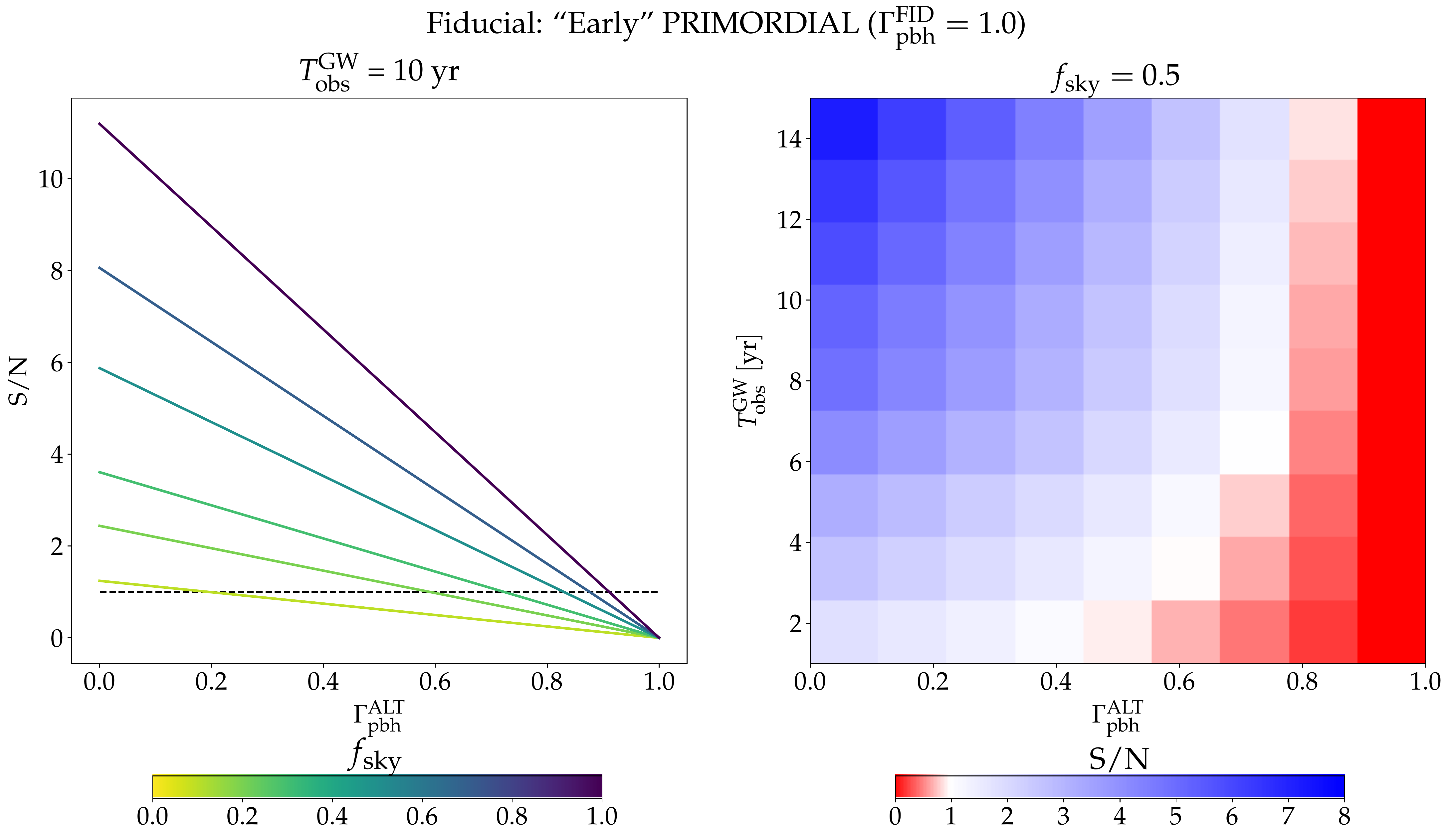}
	\caption{S/N ratios for distinguishing the Fiducial scenario from the Alternative, for different values of $T_{\mathrm{obs}}^{\rm GW}$ (from 1 yr to 15 yr), $f_{\mathrm{sky}}$ (from 0.1 to 1.0) and $\Gamma_{\mathrm{pbh}}^{\mathrm{ALT}}$ (from 0.0 to 1.0). ``Early'' primordial model assumed as Fiducial. Models with different $\Gamma_{\rm pbh}^{\rm ALT}$ assumed as Alternative. The colorbar of the right-side plots is normalized to white at $S/N=1$.}
	\label{fig:SN_primordial_fid}
\end{figure}

As figure \ref{fig:SN_primordial_fid} shows, high values of the SNR would be reached even for relatively low values of $f_{\rm sky}$ and $T_{\rm obs}^{\rm GW}$, in analogy with results from figure \ref{fig:SN_astro_fid}, in which the astrophysical scenario is instead assumed as fiducial.

\bibliography{biblio}

\providecommand{\href}[2]{#2}\begingroup\raggedright\begin{thebibliography}{100}

\bibitem{abbott:firstligodetection}
The {\bfseries LIGO Scientific Collaboration and Virgo Collaboration}, B.~P.
  Abbott {\em et~al.}, ``Observation of Gravitational Waves from a Binary Black
  Hole Merger'', \href{http://dx.doi.org/10.1103/PhysRevLett.116.061102}{{\em
  Phys. Rev. Lett.} {\bfseries 116} (Feb, 2016) 061102},
  \href{http://arxiv.org/abs/1602.03837}{{\ttfamily arXiv:1602.03837}}.

\bibitem{abbott:firstligodetectionproperties}
The {\bfseries LIGO Scientific Collaboration and Virgo Collaboration}, B.~P.
  Abbott {\em et~al.}, ``Properties of the Binary Black Hole Merger GW150914'',
  \href{http://dx.doi.org/10.1103/PhysRevLett.116.241102}{{\em Phys. Rev.
  Lett.} {\bfseries 116} (Jun, 2016) 241102},
  \href{http://arxiv.org/abs/1602.03840}{{\ttfamily arXiv:1602.03840}}.

\bibitem{Abbott:O12}
The {\bfseries LIGO Scientific Collaboration and Virgo Collaboration}, B.~P.
  Abbott {\em et~al.}, ``GWTC-1: A Gravitational-Wave Transient Catalog of
  Compact Binary Mergers Observed by LIGO and Virgo during the First and Second
  Observing Runs'', \href{http://dx.doi.org/10.1103/PhysRevX.9.031040}{{\em
  Phys. Rev. X} {\bfseries 9} (Sep, 2019) 031040}.

\bibitem{Abbott:O3}
The {\bfseries LIGO Scientific Collaboration and Virgo Collaboration},
  R.~Abbott {\em et~al.}, ``GWTC-2: Compact Binary Coalescences Observed by
  LIGO and Virgo during the First Half of the Third Observing Run'',
  \href{http://dx.doi.org/10.1103/PhysRevX.11.021053}{{\em Phys. Rev. X}
  {\bfseries 11} (Jun, 2021) 021053}.

\bibitem{kovetz17:lim}
E.~D. Kovetz, M.~P. Viero, A.~Lidz, L.~Newburgh, M.~Rahman, E.~Switzer,
  M.~Kamionkowski, J.~Aguirre, M.~Alvarez, J.~Bock, {\em et~al.},
  ``Line-intensity mapping: 2017 status report'', {\em arXiv preprint
  arXiv:1709.09066} (2017) .

\bibitem{Chang2010}
T.-C. {Chang}, U.-L. {Pen}, K.~{Bandura}, and J.~B. {Peterson}, ``{An intensity
  map of hydrogen 21-cm emission at redshift
  z\raisebox{-0.5ex}\textasciitilde0.8}'',
  \href{http://dx.doi.org/10.1038/nature09187}{{\em Nature} {\bfseries 466}
  no.~7305, (July, 2010) 463--465}.

\bibitem{Masui2013}
K.~W. {Masui}, E.~R. {Switzer}, N.~{Banavar}, K.~{Bandura}, C.~{Blake}, L.~M.
  {Calin}, T.~C. {Chang}, X.~{Chen}, Y.~C. {Li}, Y.~W. {Liao}, A.~{Natarajan},
  U.~L. {Pen}, J.~B. {Peterson}, J.~R. {Shaw}, and T.~C. {Voytek},
  ``{Measurement of 21 cm Brightness Fluctuations at z
  \raisebox{-0.5ex}\textasciitilde 0.8 in Cross-correlation}'',
  \href{http://dx.doi.org/10.1088/2041-8205/763/1/L20}{{\em The Astrophysical
  Journal Letters} {\bfseries 763} no.~1, (Jan., 2013) L20},
  \href{http://arxiv.org/abs/1208.0331}{{\ttfamily arXiv:1208.0331
  [astro-ph.CO]}}.

\bibitem{Anderson2018}
C.~J. {Anderson}, N.~J. {Luciw}, Y.~C. {Li}, C.~Y. {Kuo}, J.~{Yadav}, K.~W.
  {Masui}, T.~C. {Chang}, X.~{Chen}, N.~{Oppermann}, Y.~W. {Liao}, U.~L. {Pen},
  D.~C. {Price}, L.~{Staveley-Smith}, E.~R. {Switzer}, P.~T. {Timbie}, and
  L.~{Wolz}, ``{Low-amplitude clustering in low-redshift 21-cm intensity maps
  cross-correlated with 2dF galaxy densities}'',
  \href{http://dx.doi.org/10.1093/mnras/sty346}{{\em Monthly Notices of the
  Royal Astronomical Society} {\bfseries 476} no.~3, (May, 2018) 3382--3392},
  \href{http://arxiv.org/abs/1710.00424}{{\ttfamily arXiv:1710.00424
  [astro-ph.CO]}}.

\bibitem{Wolz2021}
L.~Wolz, A.~Pourtsidou, K.~W. Masui, T.-C. Chang, J.~E. Bautista, E.-M.
  M{\"u}ller, S.~Avila, D.~Bacon, W.~J. Percival, S.~Cunnington, {\em et~al.},
  ``HI constraints from the cross-correlation of eBOSS galaxies and Green Bank
  Telescope intensity maps'', {\em arXiv preprint arXiv:2102.04946} (2021) .

\bibitem{Villaescusa+14:reio}
F.~Villaescusa-Navarro, M.~Viel, K.~K. Datta, and T.~R. Choudhury, ``Modeling
  the neutral hydrogen distribution in the post-reionization Universe:
  intensity mapping'',
  \href{http://dx.doi.org/10.1088/1475-7516/2014/09/050}{{\em Journal of
  Cosmology and Astroparticle Physics} {\bfseries 2014} no.~09, (2014) 050}.

\bibitem{Braun2015:ska}
R.~Braun, T.~L. Bourke, J.~A. Green, E.~Keane, and J.~Wagg, ``Advancing
  astrophysics with the square kilometre array'', in {\em Advancing
  Astrophysics with the Square Kilometre Array}, vol.~215, p.~174, SISSA
  Medialab.
\newblock 2015.

\bibitem{Santos+17}
M.~G. Santos, M.~Cluver, M.~Hilton, M.~Jarvis, G.~I. Jozsa, L.~Leeuw,
  O.~Smirnov, R.~Taylor, F.~Abdalla, J.~Afonso, {\em et~al.}, ``MeerKLASS:
  MeerKAT large area synoptic survey'', {\em arXiv preprint arXiv:1709.06099}
  (2017) .

\bibitem{Wang+21:meerkat}
J.~Wang, M.~G. Santos, P.~Bull, K.~Grainge, S.~Cunnington, J.~Fonseca, M.~O.
  Irfan, Y.~Li, A.~Pourtsidou, P.~S. Soares, {\em et~al.}, ``HI intensity
  mapping with MeerKAT: Calibration pipeline for multi-dish autocorrelation
  observations'', {\em arXiv preprint arXiv:2011.13789} (2020) .

\bibitem{SKA_redbook}
D.~J. Bacon, R.~A. Battye, P.~Bull, S.~Camera, P.~G. Ferreira, I.~Harrison,
  D.~Parkinson, A.~Pourtsidou, M.~G. Santos, L.~Wolz, and et~al., ``Cosmology
  with Phase 1 of the Square Kilometre Array Red Book 2018: Technical
  specifications and performance forecasts'',
  \href{http://dx.doi.org/10.1017/pasa.2019.51}{{\em Publications of the
  Astronomical Society of Australia} {\bfseries 37} (2020) e007}.

\bibitem{maartens:ska}
R.~Maartens, F.~B. Abdalla, M.~Jarvis, and M.~G. Santos, ``Cosmology with the
  SKA -- overview'', 2015.

\bibitem{Santos15:SKA}
M.~G. Santos, P.~Bull, D.~Alonso, S.~Camera, P.~G. Ferreira, G.~Bernardi,
  R.~Maartens, M.~Viel, F.~Villaescusa-Navarro, F.~B. Abdalla, {\em et~al.},
  ``Cosmology with a SKA HI intensity mapping survey'', {\em arXiv preprint
  arXiv:1501.03989} (2015) .

\bibitem{bandura2014}
K.~{Bandura} {\em et~al.},
  \href{http://dx.doi.org/10.1117/12.2054950}{``{Canadian Hydrogen Intensity
  Mapping Experiment (CHIME) pathfinder}'',} in {\em Ground-based and Airborne
  Telescopes V}, L.~M. {Stepp}, R.~{Gilmozzi}, and H.~J. {Hall}, eds.,
  vol.~9145 of {\em Society of Photo-Optical Instrumentation Engineers (SPIE)
  Conference Series}, p.~914522.
\newblock July, 2014.
\newblock \href{http://arxiv.org/abs/1406.2288}{{\ttfamily arXiv:1406.2288
  [astro-ph.IM]}}.

\bibitem{Hu2020}
W.~{Hu}, X.~{Wang}, F.~{Wu}, Y.~{Wang}, P.~{Zhang}, and X.~{Chen}, ``{Forecast
  for FAST: from galaxies survey to intensity mapping}'',
  \href{http://dx.doi.org/10.1093/mnras/staa650}{{\em Monthly Notices of the
  Royal Astronomical Society} {\bfseries 493} no.~4, (Apr., 2020) 5854--5870},
  \href{http://arxiv.org/abs/1909.10946}{{\ttfamily arXiv:1909.10946
  [astro-ph.CO]}}.

\bibitem{Battye2016}
R.~{Battye} {\em et~al.}, ``{Update on the BINGO 21cm intensity mapping
  experiment}'', {\em arXiv e-prints} (Oct., 2016) arXiv:1610.06826,
  \href{http://arxiv.org/abs/1610.06826}{{\ttfamily arXiv:1610.06826
  [astro-ph.CO]}}.

\bibitem{Tianlai}
S.~{Das} {\em et~al.}, \href{http://dx.doi.org/10.1117/12.2313031}{``{Progress
  in the construction and testing of the Tianlai radio interferometers}'',} in
  {\em Millimeter, Submillimeter, and Far-Infrared Detectors and
  Instrumentation for Astronomy IX}, J.~{Zmuidzinas} and J.-R. {Gao}, eds.,
  vol.~10708 of {\em Society of Photo-Optical Instrumentation Engineers (SPIE)
  Conference Series}, p.~1070836.
\newblock July, 2018.
\newblock \href{http://arxiv.org/abs/1806.04698}{{\ttfamily arXiv:1806.04698
  [astro-ph.IM]}}.

\bibitem{Newburgh2016}
L.~B. {Newburgh} {\em et~al.},
  \href{http://dx.doi.org/10.1117/12.2234286}{``{HIRAX: a probe of dark energy
  and radio transients}'',} in {\em Ground-based and Airborne Telescopes VI},
  H.~J. {Hall}, R.~{Gilmozzi}, and H.~K. {Marshall}, eds., vol.~9906 of {\em
  Society of Photo-Optical Instrumentation Engineers (SPIE) Conference Series},
  p.~99065X.
\newblock Aug., 2016.
\newblock \href{http://arxiv.org/abs/1607.02059}{{\ttfamily arXiv:1607.02059
  [astro-ph.IM]}}.

\bibitem{nolta:2004}
M.~R. Nolta, E.~Wright, L.~Page, C.~Bennett, M.~Halpern, G.~Hinshaw,
  N.~Jarosik, A.~Kogut, M.~Limon, S.~Meyer, {\em et~al.}, ``First year
  Wilkinson microwave anisotropy probe observations: dark energy induced
  correlation with radio sources'',
  \href{http://dx.doi.org/10.1086/386536}{{\em The Astrophysical Journal}
  {\bfseries 608} no.~1, (2004) 10}.

\bibitem{ho:correlation}
S.~Ho, C.~Hirata, N.~Padmanabhan, U.~Seljak, and N.~Bahcall, ``Correlation of
  CMB with large-scale structure. I. Integrated Sachs-Wolfe tomography and
  cosmological implications'', {\em Physical Review D} {\bfseries 78} no.~4,
  (2008) 043519.

\bibitem{hirata:correlation}
C.~M. Hirata, S.~Ho, N.~Padmanabhan, U.~Seljak, and N.~A. Bahcall,
  ``Correlation of CMB with large-scale structure. II. Weak lensing'', {\em
  Physical Review D} {\bfseries 78} no.~4, (2008) 043520.

\bibitem{raccanelli:crosscorrelation}
A.~Raccanelli, A.~Bonaldi, M.~Negrello, S.~Matarrese, G.~Tormen, and
  G.~De~Zotti, ``A reassessment of the evidence of the Integrated Sachs-Wolfe
  effect through the WMAP-NVSS correlation'',
  \href{http://dx.doi.org/10.1111/j.1365-2966.2008.13189.x}{{\em Monthly
  Notices of the Royal Astronomical Society} {\bfseries 386} no.~4, (2008)
  2161--2166}, \href{http://arxiv.org/abs/0802.0084}{{\ttfamily
  arXiv:0802.0084}}.

\bibitem{raccanelli:radio}
A.~{Raccanelli}, G.-B. {Zhao}, D.~J. {Bacon}, M.~J. {Jarvis}, W.~J. {Percival},
  R.~P. {Norris}, H.~{R{\"o}ttgering}, F.~B. {Abdalla}, C.~M. {Cress}, J.-C.
  {Kubwimana}, S.~{Lindsay}, R.~C. {Nichol}, M.~G. {Santos}, and D.~J.
  {Schwarz}, ``{Cosmological measurements with forthcoming radio continuum
  surveys}'', \href{http://dx.doi.org/10.1111/j.1365-2966.2012.20634.x}{{\em
  Monthly Notices of the Royal Astronomical Society} {\bfseries 424} no.~2,
  (Aug., 2012) 801--819}, \href{http://arxiv.org/abs/1108.0930}{{\ttfamily
  arXiv:1108.0930 [astro-ph.CO]}}.

\bibitem{raccanelli:isw}
A.~{Raccanelli}, O.~{Dor{\'e}}, D.~J. {Bacon}, R.~{Maartens}, M.~G. {Santos},
  S.~{Camera}, T.~M. {Davis}, M.~J. {Drinkwater}, M.~{Jarvis}, R.~{Norris}, and
  D.~{Parkinson}, ``{Probing primordial non-Gaussianity via iSW measurements
  with SKA continuum surveys}'',
  \href{http://dx.doi.org/10.1088/1475-7516/2015/01/042}{{\em Journal of
  Cosmology and Astroparticle Physics} {\bfseries 2015} no.~1, (Jan., 2015)
  042}, \href{http://arxiv.org/abs/1406.0010}{{\ttfamily arXiv:1406.0010
  [astro-ph.CO]}}.

\bibitem{Bianchini:2014dla}
The {\bfseries Herschel ATLAS}, F.~Bianchini {\em et~al.}, ``{Cross-correlation
  between the CMB lensing potential measured by Planck and high-z sub-mm
  galaxies detected by the Herschel-ATLAS survey}'',
  \href{http://dx.doi.org/10.1088/0004-637X/802/1/64}{{\em Astrophys. J.}
  {\bfseries 802} no.~1, (2015) 64},
\href{http://arxiv.org/abs/1410.4502}{{\ttfamily arXiv:1410.4502
  [astro-ph.CO]}}.

\bibitem{Bianchini:2015fiw}
F.~Bianchini and A.~Lapi, ``{Cross-correlation between cosmological and
  astrophysical datasets: the Planck and Herschel case}'',
\href{http://dx.doi.org/10.1017/S1743921314013647}{{\em IAU Symp.} {\bfseries
  306} (2014) 202--205}.

\bibitem{Bianchini:2015yly}
F.~Bianchini {\em et~al.}, ``{Toward a tomographic analysis of the
  cross-correlation between Planck CMB lensing and H-ATLAS galaxies}'',
  \href{http://dx.doi.org/10.3847/0004-637X/825/1/24}{{\em Astrophys. J.}
  {\bfseries 825} no.~1, (2016) 24},
\href{http://arxiv.org/abs/1511.05116}{{\ttfamily arXiv:1511.05116
  [astro-ph.CO]}}.

\bibitem{Mukherjee:gwxcmb}
S.~Mukherjee, B.~D. Wandelt, and J.~Silk, ``Multimessenger tests of gravity
  with weakly lensed gravitational waves'',
  \href{http://dx.doi.org/10.1103/PhysRevD.101.103509}{{\em Phys. Rev. D}
  {\bfseries 101} (May, 2020) 103509}.

\bibitem{fang:cross}
K.~Fang, A.~Banerjee, E.~Charles, and Y.~Omori, ``A Cross-Correlation Study of
  High-energy Neutrinos and Tracers of Large-Scale Structure'', {\em The
  Astrophysical Journal} {\bfseries 894} no.~2, (2020) 112.

\bibitem{Martinez:cross}
H.~J. Mart{\'\i}nez, M.~E. Merch{\'a}n, C.~A. Valotto, and D.~G. Lambas,
  ``Quasar-galaxy and AGN-galaxy cross-correlations'', {\em The Astrophysical
  Journal} {\bfseries 514} no.~2, (1999) 558.

\bibitem{Jain:cross}
B.~Jain, R.~Scranton, and R.~K. Sheth, ``{Quasar—galaxy and galaxy—galaxy
  cross-correlations: model predictions with realistic galaxies}'',
  \href{http://dx.doi.org/10.1046/j.1365-8711.2003.06965.x}{{\em Monthly
  Notices of the Royal Astronomical Society} {\bfseries 345} no.~1, (10, 2003)
  62--70}.

\bibitem{Yang:cross}
X.~Yang, H.~J. Mo, F.~C. van~den Bosch, S.~M. Weinmann, C.~Li, and Y.~P. Jing,
  ``{The cross-correlation between galaxies and groups: probing the galaxy
  distribution in and around dark matter haloes}'',
  \href{http://dx.doi.org/10.1111/j.1365-2966.2005.09351.x}{{\em Monthly
  Notices of the Royal Astronomical Society} {\bfseries 362} no.~2, (09, 2005)
  711--726}.

\bibitem{Paech:cross}
K.~Paech, N.~Hamaus, B.~Hoyle, M.~Costanzi, T.~Giannantonio, S.~Hagstotz,
  G.~Sauerwein, and J.~Weller, ``Cross-correlation of galaxies and galaxy
  clusters in the Sloan Digital Sky Survey and the importance of non-Poissonian
  shot noise'', \href{http://dx.doi.org/10.1093/mnras/stx1354}{{\em Monthly
  Notices of the Royal Astronomical Society} {\bfseries 470} no.~3, (06, 2017)
  2566--2577}.

\bibitem{Oguri:2016}
M.~{Oguri}, ``{Measuring the distance-redshift relation with the
  cross-correlation of gravitational wave standard sirens and galaxies}'',
  \href{http://dx.doi.org/10.1103/PhysRevD.93.083511}{{\em Physical Review D}
  {\bfseries 93} no.~8, (Apr., 2016) 083511},
  \href{http://arxiv.org/abs/1603.02356}{{\ttfamily arXiv:1603.02356
  [astro-ph.CO]}}.

\bibitem{raccanelli:pbhprogenitors}
A.~Raccanelli, E.~D. Kovetz, S.~Bird, I.~Cholis, and J.~B. Mu\~noz,
  ``Determining the progenitors of merging black-hole binaries'',
  \href{http://dx.doi.org/10.1103/PhysRevD.94.023516}{{\em Phys. Rev. D}
  {\bfseries 94} (Jul, 2016) 023516},
  \href{http://arxiv.org/abs/1605.01405}{{\ttfamily arXiv:1605.01405}}.

\bibitem{Scelfo18:gwxlss}
G.~Scelfo, N.~Bellomo, A.~Raccanelli, S.~Matarrese, and L.~Verde,
  ``{GW}{$\times$}{LSS}: chasing the progenitors of merging binary black
  holes'', \href{http://dx.doi.org/10.1088/1475-7516/2018/09/039}{{\em JCAP}
  {\bfseries 2018} no.~09, (Sep, 2018) 039--039},
  \href{http://arxiv.org/abs/1809.03528v1}{{\ttfamily arXiv:1809.03528v1}}.

\bibitem{Scelfo20:gws}
G.~Scelfo, L.~Boco, A.~Lapi, and M.~Viel, ``Exploring galaxies-gravitational
  waves cross-correlations as an astrophysical probe'',
  \href{http://dx.doi.org/10.1088/1475-7516/2020/10/045}{{\em Journal of
  Cosmology and Astroparticle Physics} {\bfseries 2020} no.~10, (Oct, 2020)
  045--045}.

\bibitem{namikawa:cross_ng}
T.~Namikawa, A.~Nishizawa, and A.~Taruya, ``Anisotropies of gravitational-wave
  standard sirens as a new cosmological probe without redshift information'',
  {\em Physical review letters} {\bfseries 116} no.~12, (2016) 121302.

\bibitem{alonso:cross}
D.~Alonso, G.~Cusin, P.~G. Ferreira, and C.~Pitrou, ``Detecting the anisotropic
  astrophysical gravitational wave background in the presence of shot noise
  through cross-correlations'',
  \href{http://arxiv.org/abs/2002.02888}{{\ttfamily arXiv:2002.02888}}.

\bibitem{Canas:sgwb}
G.~Ca\~nas Herrera, O.~Contigiani, and V.~Vardanyan, ``Cross-correlation of the
  astrophysical gravitational-wave background with galaxy clustering'',
  \href{http://dx.doi.org/10.1103/PhysRevD.102.043513}{{\em Phys. Rev. D}
  {\bfseries 102} (Aug, 2020) 043513}.

\bibitem{Calore:crosscorrelating}
F.~Calore, A.~Cuoco, T.~Regimbau, S.~Sachdev, and P.~D. Serpico,
  ``Cross-correlating galaxy catalogs and gravitational waves: a tomographic
  approach'', \href{http://dx.doi.org/10.1103/PhysRevResearch.2.023314}{{\em
  Physical Review Research} {\bfseries 2} no.~2, (2020) 023314}.

\bibitem{camera:gwlensing}
S.~Camera and A.~Nishizawa, ``Beyond Concordance Cosmology with Magnification
  of Gravitational-Wave Standard Sirens'',
  \href{http://dx.doi.org/10.1103/PhysRevLett.110.151103}{{\em Phys. Rev.
  Lett.} {\bfseries 110} (Apr, 2013) 151103},
  \href{http://arxiv.org/abs/1303.5446}{{\ttfamily arXiv:1303.5446}}.

\bibitem{Libanore+21}
S.~Libanore, M.~C. Artale, D.~Karagiannis, M.~Liguori, N.~Bartolo,
  Y.~Bouffanais, N.~Giacobbo, M.~Mapelli, and S.~Matarrese, ``Gravitational
  Wave mergers as tracers of Large Scale Structures'',
  \href{http://dx.doi.org/10.1088/1475-7516/2021/02/035}{{\em Journal of
  Cosmology and Astroparticle Physics} {\bfseries 2021} no.~02, (Feb, 2021)
  035--035}. \url{https://doi.org/10.1088/1475-7516/2021/02/035}.

\bibitem{Mukherjee:gwxlss1}
S.~Mukherjee, B.~D. Wandelt, and J.~Silk, ``{Probing the theory of gravity with
  gravitational lensing of gravitational waves and galaxy surveys}'',
  \href{http://dx.doi.org/10.1093/mnras/staa827}{{\em Monthly Notices of the
  Royal Astronomical Society} {\bfseries 494} no.~2, (03, 2020) 1956--1970}.

\bibitem{Mukherjee:gwxlss2}
S.~Mukherjee, B.~D. Wandelt, S.~M. Nissanke, and A.~Silvestri, ``Accurate
  precision cosmology with redshift unknown gravitational wave sources'',
  \href{http://dx.doi.org/10.1103/PhysRevD.103.043520}{{\em Phys. Rev. D}
  {\bfseries 103} (Feb, 2021) 043520}.

\bibitem{Mukherjee:sgwb}
S.~Mukherjee and J.~Silk, ``{Time dependence of the astrophysical stochastic
  gravitational wave background}'',
  \href{http://dx.doi.org/10.1093/mnras/stz3226}{{\em Monthly Notices of the
  Royal Astronomical Society} {\bfseries 491} no.~4, (11, 2019) 4690--4701}.

\bibitem{canas2021gaus}
G.~Ca{\~n}as-Herrera, O.~Contigiani, and V.~Vardanyan, ``Learning how to surf:
  Reconstructing the propagation and origin of gravitational waves with
  Gaussian Processes'', {\em arXiv preprint arXiv:2105.04262} (2021) .

\bibitem{Schmidt13:cross}
S.~J. Schmidt, B.~Ménard, R.~Scranton, C.~Morrison, and C.~K. McBride,
  ``{Recovering redshift distributions with cross-correlations: pushing the
  boundaries}'', \href{http://dx.doi.org/10.1093/mnras/stt410}{{\em Monthly
  Notices of the Royal Astronomical Society} {\bfseries 431} no.~4, (04, 2013)
  3307--3318}.

\bibitem{Alonso15}
D.~Alonso and P.~G. Ferreira, ``Constraining ultralarge-scale cosmology with
  multiple tracers in optical and radio surveys'', {\em Phys. Rev. D}
  {\bfseries 92} (Sep, 2015) 063525.

\bibitem{Kovetz16:cross}
E.~D. Kovetz, A.~Raccanelli, and M.~Rahman, ``{Cosmological constraints with
  clustering-based redshifts}'',
  \href{http://dx.doi.org/10.1093/mnras/stx691}{{\em Monthly Notices of the
  Royal Astronomical Society} {\bfseries 468} no.~3, (03, 2017) 3650--3656}.

\bibitem{Alonso16:cross}
D.~Alonso and P.~G. Ferreira, ``Constraining ultralarge-scale cosmology with
  multiple tracers in optical and radio surveys'',
  \href{http://dx.doi.org/10.1103/PhysRevD.92.063525}{{\em Phys. Rev. D}
  {\bfseries 92} (Sep, 2015) 063525}.

\bibitem{Wolz2016:cross}
L.~Wolz, C.~Tonini, C.~Blake, and J.~S.~B. Wyithe, ``{Intensity mapping
  cross-correlations: connecting the largest scales to galaxy evolution}'',
  \href{http://dx.doi.org/10.1093/mnras/stw535}{{\em Monthly Notices of the
  Royal Astronomical Society} {\bfseries 458} no.~3, (03, 2016) 3399--3410}.

\bibitem{Pourtsidou16:cross}
A.~Pourtsidou, D.~Bacon, R.~Crittenden, and R.~B. Metcalf, ``{Prospects for
  clustering and lensing measurements with forthcoming intensity mapping and
  optical surveys}'', \href{http://dx.doi.org/10.1093/mnras/stw658}{{\em
  Monthly Notices of the Royal Astronomical Society} {\bfseries 459} no.~1,
  (03, 2016) 863--870}.

\bibitem{Pourtsidou16:cross_2}
A.~Pourtsidou, ``{Testing gravity at large scales with HI intensity mapping}'',
  \href{http://dx.doi.org/10.1093/mnras/stw1406}{{\em Monthly Notices of the
  Royal Astronomical Society} {\bfseries 461} no.~2, (06, 2016) 1457--1464}.

\bibitem{Raccanelli16:cross}
A.~Raccanelli, E.~Kovetz, L.~Dai, and M.~Kamionkowski, ``Detecting the
  integrated Sachs-Wolfe effect with high-redshift 21-cm surveys'',
  \href{http://dx.doi.org/10.1103/PhysRevD.93.083512}{{\em Phys. Rev. D}
  {\bfseries 93} (Apr, 2016) 083512}.
  \url{https://link.aps.org/doi/10.1103/PhysRevD.93.083512}.

\bibitem{Pourtsidou17:IM}
A.~Pourtsidou, D.~Bacon, and R.~Crittenden, ``{HI and cosmological constraints
  from intensity mapping, optical and CMB surveys}'',
  \href{http://dx.doi.org/10.1093/mnras/stx1479}{{\em Monthly Notices of the
  Royal Astronomical Society} {\bfseries 470} no.~4, (06, 2017) 4251--4260}.

\bibitem{Wolz17:cross}
L.~Wolz, C.~Blake, and J.~S.~B. Wyithe, ``{Determining the HI content of
  galaxies via intensity mapping cross-correlations}'',
  \href{http://dx.doi.org/10.1093/mnras/stx1388}{{\em Monthly Notices of the
  Royal Astronomical Society} {\bfseries 470} no.~3, (06, 2017) 3220--3226}.

\bibitem{Alonso17:lssxim}
D.~Alonso, P.~G. Ferreira, M.~J. Jarvis, and K.~Moodley, ``Calibrating
  photometric redshifts with intensity mapping observations'',
  \href{http://dx.doi.org/10.1103/PhysRevD.96.043515}{{\em Phys. Rev. D}
  {\bfseries 96} (Aug, 2017) 043515}.

\bibitem{Wolz18:cross}
L.~Wolz, S.~G. Murray, C.~Blake, and J.~S. Wyithe, ``{Intensity mapping
  cross-correlations II: HI halo models including shot noise}'',
  \href{http://dx.doi.org/10.1093/mnras/sty3142}{{\em Monthly Notices of the
  Royal Astronomical Society} {\bfseries 484} no.~1, (11, 2018) 1007--1020}.

\bibitem{Cunnington18:lssxim}
S.~Cunnington, I.~Harrison, A.~Pourtsidou, and D.~Bacon, ``{HI intensity
  mapping for clustering-based redshift estimation}'',
  \href{http://dx.doi.org/10.1093/mnras/sty2928}{{\em Monthly Notices of the
  Royal Astronomical Society} {\bfseries 482} no.~3, (10, 2018) 3341--3355}.

\bibitem{Sathyaprakash:ET}
B.~Sathyaprakash {\em et~al.}, ``Scientific objectives of Einstein Telescope'',
  \href{http://dx.doi.org/10.1088/0264-9381/29/12/124013}{{\em Classical and
  Quantum Gravity} {\bfseries 29} no.~12, (2012) 124013},
  \href{http://arxiv.org/abs/1206.0331}{{\ttfamily arXiv:1206.0331}}.

\bibitem{Newman_2008}
J.~A. Newman, ``Calibrating Redshift Distributions beyond Spectroscopic Limits
  with Cross-Correlations'', \href{http://dx.doi.org/10.1086/589982}{{\em The
  Astrophysical Journal} {\bfseries 684} no.~1, (Sep, 2008) 88--101}.

\bibitem{Benjamin_2010}
J.~Benjamin, L.~Van~Waerbeke, B.~Ménard, and M.~Kilbinger, ``{Photometric
  redshifts: estimating their contamination and distribution using clustering
  information}'',
  \href{http://dx.doi.org/10.1111/j.1365-2966.2010.17191.x}{{\em Monthly
  Notices of the Royal Astronomical Society} {\bfseries 408} no.~2, (10, 2010)
  1168--1180}.

\bibitem{Matthews_2010}
D.~J. Matthews and J.~A. Newman, ``{Reconstructing Redshift Distributions with
  Cross-Correlations: Tests and an Optimized Recipe}'',
  \href{http://dx.doi.org/10.1088/0004-637x/721/1/456}{{\em The Astrophysical
  Journal} {\bfseries 721} no.~1, (Aug, 2010) 456--468}.

\bibitem{Schmidt_2013}
S.~J. Schmidt, B.~Ménard, R.~Scranton, C.~Morrison, and C.~K. McBride,
  ``{Recovering redshift distributions with cross-correlations: pushing the
  boundaries}'', \href{http://dx.doi.org/10.1093/mnras/stt410}{{\em Monthly
  Notices of the Royal Astronomical Society} {\bfseries 431} no.~4, (04, 2013)
  3307--3318}.

\bibitem{menard2013clustering}
B.~M{\'e}nard, R.~Scranton, S.~Schmidt, C.~Morrison, D.~Jeong, T.~Budavari, and
  M.~Rahman, ``Clustering-based redshift estimation: method and application to
  data'', {\em arXiv preprint arXiv:1303.4722} (2013) .

\bibitem{McQuinn13:cbr}
M.~McQuinn and M.~White, ``{On using angular cross-correlations to determine
  source redshift distributions}'',
  \href{http://dx.doi.org/10.1093/mnras/stt914}{{\em Monthly Notices of the
  Royal Astronomical Society} {\bfseries 433} no.~4, (07, 2013) 2857--2883}.

\bibitem{Choi_2016}
A.~Choi, C.~Heymans, C.~Blake, H.~Hildebrandt, C.~A.~J. Duncan, T.~Erben,
  R.~Nakajima, L.~Van~Waerbeke, and M.~Viola, ``{CFHTLenS and RCSLenS: testing
  photometric redshift distributions using angular cross-correlations with
  spectroscopic galaxy surveys}'',
  \href{http://dx.doi.org/10.1093/mnras/stw2241}{{\em Monthly Notices of the
  Royal Astronomical Society} {\bfseries 463} no.~4, (09, 2016) 3737--3754}.

\bibitem{Scottez_2016}
V.~Scottez, Y.~Mellier, B.~R. Granett, Moutard, {\em et~al.},
  ``{Clustering-based redshift estimation: application to VIPERS/CFHTLS}'',
  \href{http://dx.doi.org/10.1093/mnras/stw1500}{{\em Monthly Notices of the
  Royal Astronomical Society} {\bfseries 462} no.~2, (07, 2016) 1683--1696}.

\bibitem{Rahman_2016}
M.~Rahman, B.~Ménard, and R.~Scranton, ``{Exploring the 2MASS extended and
  point source catalogues with clustering redshifts}'',
  \href{http://dx.doi.org/10.1093/mnras/stw256}{{\em Monthly Notices of the
  Royal Astronomical Society} {\bfseries 457} no.~4, (02, 2016) 3912--3921}.

\bibitem{Johnson_2016}
A.~Johnson, C.~Blake, A.~Amon, T.~Erben, {\em et~al.}, ``{2dFLenS and KiDS:
  determining source redshift distributions with cross-correlations}'',
  \href{http://dx.doi.org/10.1093/mnras/stw3033}{{\em Monthly Notices of the
  Royal Astronomical Society} {\bfseries 465} no.~4, (11, 2016) 4118--4132}.

\bibitem{Daalen_2018}
M.~P. van Daalen and M.~White, ``{A cross-correlation-based estimate of the
  galaxy luminosity function}'',
  \href{http://dx.doi.org/10.1093/mnras/sty545}{{\em Monthly Notices of the
  Royal Astronomical Society} {\bfseries 476} no.~4, (03, 2018) 4649--4661}.

\bibitem{Alonso_2021}
D.~Alonso, E.~Bellini, C.~Hale, M.~J. Jarvis, and D.~J. Schwarz,
  ``{Cross-correlating radio continuum surveys and CMB lensing: constraining
  redshift distributions, galaxy bias, and cosmology}'',
  \href{http://dx.doi.org/10.1093/mnras/stab046}{{\em Monthly Notices of the
  Royal Astronomical Society} {\bfseries 502} no.~1, (01, 2021) 876--887}.

\bibitem{peebles:1973}
P.~Peebles, ``Statistical analysis of catalogs of extragalactic objects. I.
  Theory'',
  \href{http://dx.doi.org/http://adsabs.harvard.edu/doi/10.1086/152431}{{\em
  The Astrophysical Journal} {\bfseries 185} (1973) 413--440}.

\bibitem{peebles:1980}
P.~PEEBLES, ``The large-scale structure of the universe(Book)'', {\em Research
  supported by the National Science Foundation. Princeton, N. J., Princeton
  University Press, 1980. 435 p} (1980) .

\bibitem{Regos:1989}
E.~Regos and A.~S. Szalay, ``Multipole expansion of the large-scale velocity
  field-Using the tensor window function'',
  \href{http://dx.doi.org/10.1086/167936}{{\em The Astrophysical Journal}
  {\bfseries 345} (1989) 627--636}.

\bibitem{Scharf:1992}
C.~Scharf, Y.~Hoffman, O.~Lahav, and D.~Lynden-Bell, ``{Spherical harmonic
  analysis of IRAS galaxies: implications for the Great Attractor and Cold Dark
  Matter}'', \href{http://dx.doi.org/10.1093/mnras/256.2.229}{{\em Monthly
  Notices of the Royal Astronomical Society} {\bfseries 256} no.~2, (05, 1992)
  229--237}.

\bibitem{Lahav:1993}
O.~Lahav, K.~Fisher, Y.~Hoffman, C.~Scharf, and S.~Zaroubi, ``Wiener
  reconstruction of galaxy surveys in spherical harmonics'',
  \href{http://dx.doi.org/10.1086/187244}{{\em arXiv preprint astro-ph/9311059}
  (1993) }.

\bibitem{Fisher:1994}
K.~B. Fisher, C.~A. Scharf, and O.~Lahav, ``{A spherical harmonic approach to
  redshift distortion and a measurement of $Omega_0$ from the 1.2-Jy IRAS
  Redshift Survey}'', \href{http://dx.doi.org/10.1093/mnras/266.1.219}{{\em
  Monthly Notices of the Royal Astronomical Society} {\bfseries 266} no.~1,
  (01, 1994) 219--226}.

\bibitem{gebhardt:2018fast}
H.~S.~G. Gebhardt and D.~Jeong, ``Fast and accurate computation of projected
  two-point functions'',
  \href{http://dx.doi.org/10.1103/PhysRevD.97.023504}{{\em Physical Review D}
  {\bfseries 97} no.~2, (2018) 023504}.

\bibitem{assassi:2017}
V.~Assassi, M.~Simonovi{\'c}, and M.~Zaldarriaga, ``Efficient evaluation of
  angular power spectra and bispectra'',
  \href{http://dx.doi.org/10.1088/1475-7516/2017/11/054}{{\em Journal of
  Cosmology and Astroparticle Physics} {\bfseries 2017} no.~11, (2017) 054}.

\bibitem{RV1}
A.~Raccanelli and Z.~Vlah, ``Unequal-time effects in the LSS correlators'',
  {\em in preparation} (2022) .

\bibitem{challinor:deltag}
A.~Challinor and A.~Lewis, ``Linear power spectrum of observed source number
  counts'', \href{http://dx.doi.org/10.1103/PhysRevD.84.043516}{{\em Phys. Rev.
  D} {\bfseries 84} (Aug, 2011) 043516},
  \href{http://arxiv.org/abs/1105.5292}{{\ttfamily arXiv:1105.5292}}.

\bibitem{bonvin:cl}
C.~Bonvin and R.~Durrer, ``What galaxy surveys really measure'',
  \href{http://dx.doi.org/10.1103/PhysRevD.84.063505}{{\em Phys. Rev. D}
  {\bfseries 84} (Sep, 2011) 063505},
  \href{http://arxiv.org/abs/1105.5280}{{\ttfamily arXiv:1105.5280}}.

\bibitem{blas:class}
D.~Blas, J.~Lesgourgues, and T.~Tram, ``The Cosmic Linear Anisotropy Solving
  System (CLASS). Part II: Approximation schemes'',
  \href{http://dx.doi.org/10.1088/1475-7516/2011/07/034}{{\em Journal of
  Cosmology and Astroparticle Physics} {\bfseries 2011} no.~07, (2011) 034},
  \href{http://arxiv.org/abs/1104.2933}{{\ttfamily arXiv:1104.2933}}.

\bibitem{didio:classgal}
E.~D. Dio, F.~Montanari, J.~Lesgourgues, and R.~Durrer, ``The CLASSgal code for
  relativistic cosmological large scale structure'',
  \href{http://dx.doi.org/10.1088/1475-7516/2013/11/044}{{\em Journal of
  Cosmology and Astroparticle Physics} {\bfseries 2013} no.~11, (2013) 044},
  \href{http://arxiv.org/abs/1307.1459}{{\ttfamily arXiv:1307.1459}}.

\bibitem{Bellomo20:multiclass}
N.~Bellomo, J.~L. Bernal, G.~Scelfo, A.~Raccanelli, and L.~Verde, ``Beware of
  commonly used approximations. Part I. Errors in forecasts'',
  \href{http://dx.doi.org/10.1088/1475-7516/2020/10/016}{{\em Journal of
  Cosmology and Astroparticle Physics} {\bfseries 2020} no.~10, (Oct, 2020)
  016--016}.

\bibitem{Bernal20:multiclass}
J.~L. Bernal, N.~Bellomo, A.~Raccanelli, and L.~Verde, ``Beware of commonly
  used approximations. Part {II}. Estimating systematic biases in the best-fit
  parameters'', \href{http://dx.doi.org/10.1088/1475-7516/2020/10/017}{{\em
  Journal of Cosmology and Astroparticle Physics} {\bfseries 2020} no.~10,
  (Oct, 2020) 017--017}.

\bibitem{Kaiser:bias}
N.~Kaiser, ``On the spatial correlations of Abell clusters'',
  \href{http://dx.doi.org/10.1086/184341}{{\em The Astrophysical Journal}
  {\bfseries 284} (1984) L9--L12}.

\bibitem{Bardeen:bias}
J.~M. Bardeen, J.~Bond, N.~Kaiser, and A.~Szalay, ``The statistics of peaks of
  Gaussian random fields'', \href{http://dx.doi.org/10.1086/164143}{{\em The
  Astrophysical Journal} {\bfseries 304} (1986) 15--61}.

\bibitem{Mo:smallhalosbias}
H.~J. Mo and S.~D.~M. White, ``An analytic model for the spatial clustering of
  dark matter haloes'', \href{http://dx.doi.org/10.1093/mnras/282.2.347}{{\em
  Monthly Notices of the Royal Astronomical Society} {\bfseries 282} no.~2,
  (1996) 347--361}, \href{http://arxiv.org/abs/astro-ph/9512127}{{\ttfamily
  arXiv:astro-ph/9512127}}.

\bibitem{matarrese:clusteringevolution}
S.~Matarrese, P.~Coles, F.~Lucchin, and L.~Moscardini, ``Redshift evolution of
  clustering'', \href{http://dx.doi.org/10.1093/mnras/286.1.115}{{\em Monthly
  Notices of the Royal Astronomical Society} {\bfseries 286} no.~1, (03, 1997)
  115--132}, \href{http://arxiv.org/abs/astro-ph/9608004}{{\ttfamily
  arXiv:astro-ph/9608004}}.

\bibitem{dekel:stochasticbiasing}
A.~Dekel and O.~Lahav, ``Stochastic Nonlinear Galaxy Biasing'',
  \href{http://dx.doi.org/10.1086/307428}{{\em The Astrophysical Journal}
  {\bfseries 520} no.~1, (Jul, 1999) 24--34},
  \href{http://arxiv.org/abs/astro-ph/9806193}{{\ttfamily
  arXiv:astro-ph/9806193}}.

\bibitem{benson:galaxybias}
A.~J. Benson, S.~Cole, C.~S. Frenk, C.~M. Baugh, and C.~G. Lacey, ``The nature
  of galaxy bias and clustering'',
  \href{http://dx.doi.org/10.1046/j.1365-8711.2000.03101.x}{{\em Monthly
  Notices of the Royal Astronomical Society} {\bfseries 311} no.~4, (02, 2000)
  793--808}, \href{http://arxiv.org/abs/astro-ph/9903343}{{\ttfamily
  arXiv:astro-ph/9903343}}.

\bibitem{peacock:halooccupation}
J.~A. Peacock and R.~E. Smith, ``Halo occupation numbers and galaxy bias'',
  \href{http://dx.doi.org/10.1046/j.1365-8711.2000.03779.x}{{\em Monthly
  Notices of the Royal Astronomical Society} {\bfseries 318} no.~4, (11, 2000)
  1144--1156}, \href{http://arxiv.org/abs/astro-ph/0005010}{{\ttfamily
  arXiv:astro-ph/0005010}}.

\bibitem{Desjacques:bias}
V.~Desjacques, D.~Jeong, and F.~Schmidt, ``Large-scale galaxy bias'',
  \href{http://dx.doi.org/https://doi.org/10.1016/j.physrep.2017.12.002}{{\em
  Physics Reports} {\bfseries 733} (2018) 1 -- 193}.

\bibitem{turner:magnificationbias}
E.~L. Turner, J.~P. Ostriker, and J.~R. Gott, III, ``The statistics of
  gravitational lenses - The distributions of image angular separations and
  lens redshifts'', \href{http://dx.doi.org/10.1086/162379}{{\em Astrophysical
  Journal} {\bfseries 284} (Sep, 1984) 1--22}.

\bibitem{challinor:evolutionbias}
A.~Challinor and A.~Lewis, ``Linear power spectrum of observed source number
  counts'', \href{http://dx.doi.org/10.1103/PhysRevD.84.043516}{{\em Phys. Rev.
  D} {\bfseries 84} (Aug, 2011) 043516},
  \href{http://arxiv.org/abs/1105.5292}{{\ttfamily arXiv:1105.5292}}.

\bibitem{jeong:evolutionbias}
D.~Jeong, F.~Schmidt, and C.~M. Hirata, ``Large-scale clustering of galaxies in
  general relativity'',
  \href{http://dx.doi.org/10.1103/PhysRevD.85.023504}{{\em Phys. Rev. D}
  {\bfseries 85} (Jan, 2012) 023504},
  \href{http://arxiv.org/abs/1107.5427}{{\ttfamily arXiv:1107.5427}}.

\bibitem{bertacca:evolutionbias}
D.~Bertacca, R.~Maartens, A.~Raccanelli, and C.~Clarkson, ``Beyond the
  plane-parallel and Newtonian approach: wide-angle redshift distortions and
  convergence in general relativity'',
  \href{http://dx.doi.org/10.1088/1475-7516/2012/10/025}{{\em Journal of
  Cosmology and Astroparticle Physics} {\bfseries 2012} no.~10, (2012) 025},
  \href{http://arxiv.org/abs/1205.5221}{{\ttfamily arXiv:1205.5221}}.

\bibitem{CE:2019}
D.~Reitze, R.~X. Adhikari, S.~Ballmer, B.~Barish, L.~Barsotti, G.~Billingsley,
  D.~A. Brown, Y.~Chen, D.~Coyne, R.~Eisenstein, {\em et~al.}, ``Cosmic
  explorer: the US contribution to gravitational-wave astronomy beyond LIGO'',
  {\em arXiv preprint arXiv:1907.04833} (2019) .

\bibitem{Boco19:gws}
L.~{Boco}, A.~{Lapi}, S.~{Goswami}, F.~{Perrotta}, C.~{Baccigalupi}, and
  L.~{Danese}, ``{Merging Rates of Compact Binaries in Galaxies: Perspectives
  for Gravitational Wave Detections}'',
  \href{http://dx.doi.org/10.3847/1538-4357/ab328e}{{\em The Astrophysical
  Journal} {\bfseries 881} no.~2, (Aug., 2019) 157},
  \href{http://arxiv.org/abs/1907.06841}{{\ttfamily arXiv:1907.06841
  [astro-ph.GA]}}.

\bibitem{dominik+13}
M.~{Dominik}, K.~{Belczynski}, C.~{Fryer}, D.~E. {Holz}, E.~{Berti},
  T.~{Bulik}, I.~{Mand el}, and R.~{O'Shaughnessy}, ``{Double Compact Objects.
  II. Cosmological Merger Rates}'',
  \href{http://dx.doi.org/10.1088/0004-637X/779/1/72}{{\em The Astrophysical
  Journal} {\bfseries 779} no.~1, (Dec., 2013) 72},
  \href{http://arxiv.org/abs/1308.1546}{{\ttfamily arXiv:1308.1546
  [astro-ph.HE]}}.

\bibitem{dominik+15}
M.~{Dominik}, E.~{Berti}, R.~{O'Shaughnessy}, I.~{Mandel}, K.~{Belczynski},
  C.~{Fryer}, D.~E. {Holz}, T.~{Bulik}, and F.~{Pannarale}, ``{Double Compact
  Objects III: Gravitational-wave Detection Rates}'',
  \href{http://dx.doi.org/10.1088/0004-637X/806/2/263}{{\em The Astrophysical
  Journal} {\bfseries 806} no.~2, (June, 2015) 263},
  \href{http://arxiv.org/abs/1405.7016}{{\ttfamily arXiv:1405.7016
  [astro-ph.HE]}}.

\bibitem{demink+13}
S.~E. {de Mink}, N.~{Langer}, R.~G. {Izzard}, H.~{Sana}, and A.~{de Koter},
  ``{The Rotation Rates of Massive Stars: The Role of Binary Interaction
  through Tides, Mass Transfer, and Mergers}'',
  \href{http://dx.doi.org/10.1088/0004-637X/764/2/166}{{\em The Astrophysical
  Journal} {\bfseries 764} no.~2, (Feb., 2013) 166},
  \href{http://arxiv.org/abs/1211.3742}{{\ttfamily arXiv:1211.3742
  [astro-ph.SR]}}.

\bibitem{spera+15}
M.~{Spera}, M.~{Mapelli}, and A.~{Bressan}, ``{The mass spectrum of compact
  remnants from the PARSEC stellar evolution tracks}'',
  \href{http://dx.doi.org/10.1093/mnras/stv1161}{{\em Monthly Notices of the
  Royal Astronomical Society} {\bfseries 451} no.~4, (Aug., 2015) 4086--4103},
  \href{http://arxiv.org/abs/1505.05201}{{\ttfamily arXiv:1505.05201
  [astro-ph.SR]}}.

\bibitem{spera+17}
M.~{Spera} and M.~{Mapelli}, ``{Very massive stars, pair-instability supernovae
  and intermediate-mass black holes with the sevn code}'',
  \href{http://dx.doi.org/10.1093/mnras/stx1576}{{\em Monthly Notices of the
  Royal Astronomical Society} {\bfseries 470} no.~4, (Oct., 2017) 4739--4749},
  \href{http://arxiv.org/abs/1706.06109}{{\ttfamily arXiv:1706.06109
  [astro-ph.SR]}}.

\bibitem{spera+19}
M.~{Spera}, M.~{Mapelli}, N.~{Giacobbo}, A.~A. {Trani}, A.~{Bressan}, and
  G.~{Costa}, ``{Merging black hole binaries with the SEVN code}'',
  \href{http://dx.doi.org/10.1093/mnras/stz359}{{\em Monthly Notices of the
  Royal Astronomical Society} {\bfseries 485} no.~1, (May, 2019) 889--907},
  \href{http://arxiv.org/abs/1809.04605}{{\ttfamily arXiv:1809.04605
  [astro-ph.HE]}}.

\bibitem{giacobbo+18}
N.~{Giacobbo} and M.~{Mapelli}, ``{The progenitors of compact-object binaries:
  impact of metallicity, common envelope and natal kicks}'',
  \href{http://dx.doi.org/10.1093/mnras/sty1999}{{\em Monthly Notices of the
  Royal Astronomical Society} {\bfseries 480} no.~2, (Oct., 2018) 2011--2030},
  \href{http://arxiv.org/abs/1806.00001}{{\ttfamily arXiv:1806.00001
  [astro-ph.HE]}}.

\bibitem{belczynski:massivebhsmergers}
K.~Belczynski, M.~Dominik, T.~Bulik, R.~O'Shaughnessy, C.~Fryer, and D.~E.
  Holz, ``The Effect of Metallicity on the Detection Prospects for
  Gravitational Waves'',
  \href{http://dx.doi.org/10.1088/2041-8205/715/2/L138}{{\em The Astrophysical
  Journal Letters} {\bfseries 715} no.~2, (2010) L138},
  \href{http://arxiv.org/abs/1004.0386}{{\ttfamily arXiv:1004.0386}}.

\bibitem{lamberts+16}
A.~{Lamberts}, S.~{Garrison-Kimmel}, D.~R. {Clausen}, and P.~F. {Hopkins},
  ``{When and where did GW150914 form?}'',
  \href{http://dx.doi.org/10.1093/mnrasl/slw152}{{\em Monthly Notices of the
  Royal Astronomical Society} {\bfseries 463} no.~1, (Nov., 2016) L31--L35},
  \href{http://arxiv.org/abs/1605.08783}{{\ttfamily arXiv:1605.08783
  [astro-ph.HE]}}.

\bibitem{cao+18}
L.~{Cao}, Y.~{Lu}, and Y.~{Zhao}, ``{Host galaxy properties of mergers of
  stellar binary black holes and their implications for advanced LIGO
  gravitational wave sources}'',
  \href{http://dx.doi.org/10.1093/mnras/stx3087}{{\em Monthly Notices of the
  Royal Astronomical Society} {\bfseries 474} no.~4, (Mar., 2018) 4997--5007},
  \href{http://arxiv.org/abs/1711.09190}{{\ttfamily arXiv:1711.09190
  [astro-ph.GA]}}.

\bibitem{elbert+18}
O.~D. {Elbert}, J.~S. {Bullock}, and M.~{Kaplinghat}, ``{Counting black holes:
  The cosmic stellar remnant population and implications for LIGO}'',
  \href{http://dx.doi.org/10.1093/mnras/stx1959}{{\em Monthly Notices of the
  Royal Astronomical Society} {\bfseries 473} no.~1, (Jan., 2018) 1186--1194},
  \href{http://arxiv.org/abs/1703.02551}{{\ttfamily arXiv:1703.02551
  [astro-ph.GA]}}.

\bibitem{li+18}
S.-S. {Li}, S.~{Mao}, Y.~{Zhao}, and Y.~{Lu}, ``{Gravitational lensing of
  gravitational waves: a statistical perspective}'',
  \href{http://dx.doi.org/10.1093/mnras/sty411}{{\em Monthly Notices of the
  Royal Astronomical Society} {\bfseries 476} no.~2, (May, 2018) 2220--2229},
  \href{http://arxiv.org/abs/1802.05089}{{\ttfamily arXiv:1802.05089
  [astro-ph.CO]}}.

\bibitem{neijssel+19}
C.~J. {Neijssel}, A.~{Vigna-G{\'o}mez}, S.~{Stevenson}, J.~W. {Barrett}, S.~M.
  {Gaebel}, F.~S. {Broekgaarden}, S.~E. {de Mink}, D.~{Sz{\'e}csi},
  S.~{Vinciguerra}, and I.~{Mandel}, ``{The effect of the metallicity-specific
  star formation history on double compact object mergers}'',
  \href{http://dx.doi.org/10.1093/mnras/stz2840}{{\em Monthly Notices of the
  Royal Astronomical Society} {\bfseries 490} no.~3, (Dec., 2019) 3740--3759},
  \href{http://arxiv.org/abs/1906.08136}{{\ttfamily arXiv:1906.08136
  [astro-ph.SR]}}.

\bibitem{Abbott:O3run}
R.~Abbott, T.~Abbott, S.~Abraham, F.~Acernese, K.~Ackley, A.~Adams, C.~Adams,
  R.~Adhikari, V.~Adya, C.~Affeldt, {\em et~al.}, ``Population Properties of
  Compact Objects from the Second LIGO-Virgo Gravitational-Wave Transient
  Catalog'', {\em arXiv preprint arXiv:2010.14533} (2020) .

\bibitem{aversa+15}
R.~{Aversa}, A.~{Lapi}, G.~{de Zotti}, F.~{Shankar}, and L.~{Danese}, ``{Black
  Hole and Galaxy Coevolution from Continuity Equation and Abundance
  Matching}'', \href{http://dx.doi.org/10.1088/0004-637X/810/1/74}{{\em The
  Astrophysical Journal} {\bfseries 810} no.~1, (Sept., 2015) 74},
  \href{http://arxiv.org/abs/1507.07318}{{\ttfamily arXiv:1507.07318
  [astro-ph.GA]}}.

\bibitem{Dewdney:ska}
P.~E. {Dewdney}, P.~J. {Hall}, R.~T. {Schilizzi}, and T.~J. L.~W. {Lazio},
  ``The Square Kilometre Array'',
  \href{http://dx.doi.org/10.1109/JPROC.2009.2021005}{{\em Proceedings of the
  IEEE} {\bfseries 97} no.~8, (2009) 1482--1496}.

\bibitem{Hall13}
A.~Hall, C.~Bonvin, and A.~Challinor, ``Testing general relativity with 21-cm
  intensity mapping'', {\em Phys. Rev. D} {\bfseries 87} (Mar, 2013) 064026.

\bibitem{Crighton15:HI}
N.~H.~M. Crighton, M.~T. Murphy, J.~X. Prochaska, G.~Worseck, M.~Rafelski,
  G.~D. Becker, S.~L. Ellison, M.~Fumagalli, S.~Lopez, A.~Meiksin, and J.~M.
  O'Meara, ``{The neutral hydrogen cosmological mass density at z = 5}'',
  \href{http://dx.doi.org/10.1093/mnras/stv1182}{{\em Monthly Notices of the
  Royal Astronomical Society} {\bfseries 452} no.~1, (07, 2015) 217--234}.

\bibitem{Battye:T_HI}
R.~A. Battye, I.~W.~A. Browne, C.~Dickinson, G.~Heron, B.~Maffei, and
  A.~Pourtsidou, ``{HI intensity mapping: a single dish approach}'',
  \href{http://dx.doi.org/10.1093/mnras/stt1082}{{\em Monthly Notices of the
  Royal Astronomical Society} {\bfseries 434} no.~2, (07, 2013) 1239--1256}.

\bibitem{Spinelli20:HI}
M.~Spinelli, A.~Zoldan, G.~De Lucia, L.~Xie, and M.~Viel, ``{The atomic
  hydrogen content of the post-reionization Universe}'',
  \href{http://dx.doi.org/10.1093/mnras/staa604}{{\em Monthly Notices of the
  Royal Astronomical Society} {\bfseries 493} no.~4, (03, 2020) 5434--5455}.

\bibitem{Villaescusa+18:ingr}
F.~Villaescusa-Navarro, S.~Genel, E.~Castorina, A.~Obuljen, D.~N. Spergel,
  L.~Hernquist, D.~Nelson, I.~P. Carucci, A.~Pillepich, F.~Marinacci, {\em
  et~al.}, ``Ingredients for 21 cm intensity mapping'',
  \href{http://dx.doi.org/10.3847/1538-4357/aadba0}{{\em The Astrophysical
  Journal} {\bfseries 866} no.~2, (2018) 135}.

\bibitem{Martin_2010}
A.~M. Martin, E.~Papastergis, R.~Giovanelli, M.~P. Haynes, C.~M. Springob, and
  S.~Stierwalt, ``The Arecibo Legacy Fast ALFA Survey: X. The HI Mass Function
  and $\Omega_{\rm HI}$ From the 40\% ALFALFA Survey'',
  \href{http://dx.doi.org/10.1088/0004-637x/723/2/1359}{{\em The Astrophysical
  Journal} {\bfseries 723} no.~2, (Oct, 2010) 1359--1374}.
  \url{https://doi.org/10.1088/0004-637x/723/2/1359}.

\bibitem{Castorina+17:HI}
E.~Castorina and F.~Villaescusa-Navarro, ``{On the spatial distribution of
  neutral hydrogen in the Universe: bias and shot-noise of the HI power
  spectrum}'', \href{http://dx.doi.org/10.1093/mnras/stx1599}{{\em Monthly
  Notices of the Royal Astronomical Society} {\bfseries 471} no.~2, (06, 2017)
  1788--1796}.

\bibitem{Bull15:IM}
P.~Bull, P.~G. Ferreira, P.~Patel, and M.~G. Santos, ``Late-time cosmology with
  21 cm intensity mapping experiments'',
  \href{http://dx.doi.org/https://iopscience.iop.org/article/10.1088/0004-637X/803/1/21}{{\em
  The Astrophysical Journal} {\bfseries 803} no.~1, (2015) 21}.

\bibitem{Switzer+2013}
E.~R. Switzer, K.~W. Masui, K.~Bandura, {\em et~al.}, ``{Determination of
  z$\sim$0.8 neutral hydrogen fluctuations using the 21 cm intensity mapping
  autocorrelation}'', \href{http://dx.doi.org/10.1093/mnrasl/slt074}{{\em
  Monthly Notices of the Royal Astronomical Society: Letters} {\bfseries 434}
  no.~1, (06, 2013) L46--L50}.

\bibitem{Wolz+21}
L.~Wolz, A.~Pourtsidou, K.~W. Masui, T.-C. Chang, J.~E. Bautista, E.-M.
  M{\"u}ller, S.~Avila, D.~Bacon, W.~J. Percival, S.~Cunnington, {\em et~al.},
  ``HI constraints from the cross-correlation of eBOSS galaxies and Green Bank
  Telescope intensity maps'', {\em arXiv preprint arXiv:2102.04946} (2021) .

\bibitem{Alonso14:fogs}
D.~Alonso, P.~Bull, P.~G. Ferreira, and M.~G. Santos, ``{Blind foreground
  subtraction for intensity mapping experiments}'',
  \href{http://dx.doi.org/10.1093/mnras/stu2474}{{\em Monthly Notices of the
  Royal Astronomical Society} {\bfseries 447} no.~1, (12, 2014) 400--416}.

\bibitem{Carucci:2020enz}
I.~P. {Carucci}, M.~O. {Irfan}, and J.~{Bobin}, ``{Recovery of 21 cm intensity
  maps with sparse component separation}'',
  \href{http://dx.doi.org/10.1093/mnras/staa2854}{{\em Monthly Notices of the
  Royal Astronomical Society} (Sept., 2020) },
  \href{http://arxiv.org/abs/2006.05996}{{\ttfamily arXiv:2006.05996
  [astro-ph.CO]}}.

\bibitem{Cunnington:2020njn}
S.~{Cunnington}, M.~O. {Irfan}, I.~P. {Carucci}, A.~{Pourtsidou}, and
  J.~{Bobin}, ``{21-cm foregrounds and polarization leakage: cleaning and
  mitigation strategies}'', \href{http://dx.doi.org/10.1093/mnras/stab856}{{\em
  Monthly Notices of the Royal Astronomical Society} {\bfseries 504} no.~1,
  (June, 2021) 208--227}, \href{http://arxiv.org/abs/2010.02907}{{\ttfamily
  arXiv:2010.02907 [astro-ph.CO]}}.

\bibitem{Matshawule2020}
S.~D. {Matshawule}, M.~{Spinelli}, M.~G. {Santos}, and S.~{Ngobese}, ``{Hi
  intensity mapping with MeerKAT: Primary beam effects on foreground
  cleaning}'', {\em arXiv e-prints} (Nov., 2020) arXiv:2011.10815,
  \href{http://arxiv.org/abs/2011.10815}{{\ttfamily arXiv:2011.10815
  [astro-ph.CO]}}.

\bibitem{Soares_2021}
P.~S. {Soares}, C.~A. {Watkinson}, S.~{Cunnington}, and A.~{Pourtsidou},
  ``{Gaussian Process Regression for foreground removal in HI intensity mapping
  experiments}'', {\em arXiv e-prints} (May, 2021) arXiv:2105.12665,
  \href{http://arxiv.org/abs/2105.12665}{{\ttfamily arXiv:2105.12665
  [astro-ph.CO]}}.

\bibitem{Camera16:sys}
S.~Camera, I.~Harrison, A.~Bonaldi, and M.~L. Brown, ``{SKA weak lensing –
  III. Added value of multiwavelength synergies for the mitigation of
  systematics}'', \href{http://dx.doi.org/10.1093/mnras/stw2688}{{\em Monthly
  Notices of the Royal Astronomical Society} {\bfseries 464} no.~4, (10, 2016)
  4747--4760}.

\bibitem{Spinelli2021}
M.~{Spinelli}, I.~P. {Carucci}, S.~{Cunnington}, S.~E. {Harper}, M.~O. {Irfan},
  J.~{Fonseca}, A.~{Pourtsidou}, and L.~{Wolz}, ``{SKAO HI Intensity Mapping:
  Blind Foreground Subtraction Challenge}'', {\em arXiv e-prints} (July, 2021)
  arXiv:2107.10814, \href{http://arxiv.org/abs/2107.10814}{{\ttfamily
  arXiv:2107.10814 [astro-ph.CO]}}.

\bibitem{Schutz:1986}
B.~F. Schutz, ``Determining the Hubble constant from gravitational wave
  observations'', {\em Nature} {\bfseries 323} no.~6086, (1986) 310--311.

\bibitem{Bertacca:GWDL}
D.~Bertacca, A.~Raccanelli, N.~Bartolo, and S.~Matarrese, ``Cosmological
  perturbation effects on gravitational-wave luminosity distance estimates'',
  \href{http://dx.doi.org/https://doi.org/10.1016/j.dark.2018.03.001}{{\em
  Physics of the Dark Universe} {\bfseries 20} (2018) 32--40}.

\bibitem{Mukherjee:gwgr}
S.~Mukherjee, B.~D. Wandelt, and J.~Silk, ``{Testing the general theory of
  relativity using gravitational wave propagation from dark standard sirens}'',
  \href{http://dx.doi.org/10.1093/mnras/stab001}{{\em Monthly Notices of the
  Royal Astronomical Society} {\bfseries 502} no.~1, (01, 2021) 1136--1144}.

\bibitem{Planck:XIII}
{Planck Collaboration}, {Ade, P. A. R.}, {\em et~al.}, ``Planck 2015 results -
  XIII. Cosmological parameters'',
  \href{http://dx.doi.org/10.1051/0004-6361/201525830}{{\em A\&A} {\bfseries
  594} (2016) A13}, \href{http://arxiv.org/abs/1502.01589}{{\ttfamily
  arXiv:1502.01589}}.

\bibitem{Ng_2021}
K.~K.~Y. Ng, S.~Vitale, W.~M. Farr, and C.~L. Rodriguez, ``Probing Multiple
  Populations of Compact Binaries with Third-generation Gravitational-wave
  Detectors'', \href{http://dx.doi.org/10.3847/2041-8213/abf8be}{{\em The
  Astrophysical Journal Letters} {\bfseries 913} no.~1, (May, 2021) L5}.

\bibitem{planck:2018}
N.~Aghanim, Y.~Akrami, M.~Ashdown, J.~Aumont, C.~Baccigalupi, M.~Ballardini,
  A.~Banday, R.~Barreiro, N.~Bartolo, S.~Basak, {\em et~al.}, ``Planck 2018
  results-VI. Cosmological parameters'', {\em Astronomy \& Astrophysics}
  {\bfseries 641} (2020) A6.

\bibitem{boss:de}
S.~Alam {\em et~al.}, ``{The clustering of galaxies in the completed SDSS-III
  Baryon Oscillation Spectroscopic Survey: cosmological analysis of the DR12
  galaxy sample}'', \href{http://dx.doi.org/10.1093/mnras/stx721}{{\em Monthly
  Notices of the Royal Astronomical Society} {\bfseries 470} no.~3, (03, 2017)
  2617--2652}.

\bibitem{Giudice:2021}
G.~F. Giudice, M.~McCullough, and T.~You, ``{Self-Organised Localisation}'',
  \href{http://arxiv.org/abs/2105.08617}{{\ttfamily arXiv:2105.08617
  [hep-ph]}}.

\bibitem{Heisenberg:swampland}
L.~Heisenberg, M.~Bartelmann, R.~Brandenberger, and A.~Refregier, ``Dark energy
  in the swampland'', \href{http://dx.doi.org/10.1103/PhysRevD.98.123502}{{\em
  Phys. Rev. D} {\bfseries 98} (Dec, 2018) 123502}.

\bibitem{Raccanelli16:radio}
A.~Raccanelli, ``{Gravitational wave astronomy with radio galaxy surveys}'',
  \href{http://dx.doi.org/10.1093/mnras/stx835}{{\em Monthly Notices of the
  Royal Astronomical Society} {\bfseries 469} no.~1, (04, 2017) 656--670}.

\bibitem{blanchard2020euclid}
A.~Blanchard, S.~Camera, C.~Carbone, V.~Cardone, S.~Casas, S.~Clesse,
  S.~Ili{\'c}, M.~Kilbinger, T.~Kitching, M.~Kunz, {\em et~al.}, ``Euclid
  preparation-VII. Forecast validation for Euclid cosmological probes'',
  \href{http://dx.doi.org/10.1051/0004-6361/202038071}{{\em Astronomy \&
  Astrophysics} {\bfseries 642} (2020) A191}.

\bibitem{LSST_DE}
{\v{Z}}.~Ivezi{\'{c}} {\em et~al.}, ``{LSST}: From Science Drivers to Reference
  Design and Anticipated Data Products'',
  \href{http://dx.doi.org/10.3847/1538-4357/ab042c}{{\em The Astrophysical
  Journal} {\bfseries 873} no.~2, (Mar, 2019) 111}.

\bibitem{Bonaldi:sys}
A.~Bonaldi, I.~Harrison, S.~Camera, and M.~L. Brown, ``{SKA weak lensing– II.
  Simulated performance and survey design considerations}'',
  \href{http://dx.doi.org/10.1093/mnras/stw2104}{{\em Monthly Notices of the
  Royal Astronomical Society} {\bfseries 463} no.~4, (08, 2016) 3686--3698}.

\bibitem{hawking:pbh}
S.~Hawking, ``Gravitationally collapsed objects of very low mass'', {\em
  Monthly Notices of the Royal Astronomical Society} {\bfseries 152} no.~1,
  (1971) 75--78.

\bibitem{carr:pbh}
B.~J. Carr and S.~W. Hawking, ``Black holes in the early Universe'', {\em
  Monthly Notices of the Royal Astronomical Society} {\bfseries 168} no.~2,
  (1974) 399--415.

\bibitem{Polnarev88}
A.~Polnarev and R.~Zembowicz, ``Formation of primordial black holes by cosmic
  strings'', \href{http://dx.doi.org/10.1103/PhysRevD.43.1106}{{\em Phys. Rev.
  D} {\bfseries 43} (Feb, 1991) 1106--1109}.

\bibitem{HAWKING1989237}
S.~Hawking, ``Black holes from cosmic strings'',
  \href{http://dx.doi.org/https://doi.org/10.1016/0370-2693(89)90206-2}{{\em
  Physics Letters B} {\bfseries 231} no.~3, (1989) 237 -- 239}.

\bibitem{WICHOSKI1998191}
U.~F. Wichoski, J.~H. MacGibbon, and R.~H. Brandenberger, ``Astrophysical
  constraints on primordial black hole formation from collapsing cosmic
  strings'', \href{http://dx.doi.org/10.1016/S0370-1573(98)00070-2}{{\em
  Physics Reports} {\bfseries 307} no.~1, (1998) 191 -- 196}.

\bibitem{BEREZIN198391}
V.~Berezin, V.~Kuzmin, and I.~Tkachev, ``Thin-wall vacuum domain evolution'',
  \href{http://dx.doi.org/https://doi.org/10.1016/0370-2693(83)90630-5}{{\em
  Physics Letters B} {\bfseries 120} no.~1, (1983) 91 -- 96}.

\bibitem{Ipser84}
J.~Ipser and P.~Sikivie, ``Gravitationally repulsive domain wall'',
  \href{http://dx.doi.org/10.1103/PhysRevD.30.712}{{\em Phys. Rev. D}
  {\bfseries 30} (Aug, 1984) 712--719}.

\bibitem{ivanov:pbhfrominflationI}
P.~Ivanov, P.~Naselsky, and I.~Novikov, ``Inflation and primordial black holes
  as dark matter'', \href{http://dx.doi.org/10.1103/PhysRevD.50.7173}{{\em
  Phys. Rev. D} {\bfseries 50} (Dec, 1994) 7173--7178}.

\bibitem{Bellido:pbh}
J.~Garc\'{\i}a-Bellido, A.~Linde, and D.~Wands, ``Density perturbations and
  black hole formation in hybrid inflation'',
  \href{http://dx.doi.org/10.1103/PhysRevD.54.6040}{{\em Phys. Rev. D}
  {\bfseries 54} (Nov, 1996) 6040--6058},
  \href{http://arxiv.org/abs/astro-ph/9605094}{{\ttfamily
  arXiv:astro-ph/9605094}}.

\bibitem{ivanov:pbhfrominflationII}
P.~Ivanov, ``Nonlinear metric perturbations and production of primordial black
  holes'', \href{http://dx.doi.org/10.1103/PhysRevD.57.7145}{{\em Phys. Rev. D}
  {\bfseries 57} (Jun, 1998) 7145--7154},
  \href{http://arxiv.org/abs/astro-ph/9708224}{{\ttfamily
  arXiv:astro-ph/9708224}}.

\bibitem{Crawford82}
M.~Crawford and D.~N. Schramm, ``Spontaneous generation of density
  perturbations in the early Universe'',
  \href{http://dx.doi.org/http://dx.doi.org/10.1038/298538a0}{{\em Nature}
  {\bfseries 298} (1982) 538--540}.

\bibitem{LA1989375}
D.~La and P.~J. Steinhardt, ``Bubble percolation in extended inflationary
  models'',
  \href{http://dx.doi.org/https://doi.org/10.1016/0370-2693(89)90890-3}{{\em
  Physics Letters B} {\bfseries 220} no.~3, (1989) 375 -- 378}.

\bibitem{musco:2005}
I.~Musco, J.~C. Miller, and L.~Rezzolla, ``Computations of primordial
  black-hole formation'',
  \href{http://dx.doi.org/10.1088/0264-9381/22/7/013}{{\em Classical and
  Quantum Gravity} {\bfseries 22} no.~7, (Mar, 2005) 1405--1424}.

\bibitem{cole:2018}
P.~S. Cole and C.~T. Byrnes, ``Extreme scenarios: the tightest possible
  constraints on the power spectrum due to primordial black holes'',
  \href{http://dx.doi.org/10.1088/1475-7516/2018/02/019}{{\em Journal of
  Cosmology and Astroparticle Physics} {\bfseries 2018} no.~02, (Feb, 2018)
  019--019}.

\bibitem{kalaja:2019}
A.~Kalaja, N.~Bellomo, N.~Bartolo, D.~Bertacca, S.~Matarrese, I.~Musco,
  A.~Raccanelli, and L.~Verde, ``From primordial black holes abundance to
  primordial curvature power spectrum (and back)'',
  \href{http://dx.doi.org/10.1088/1475-7516/2019/10/031}{{\em Journal of
  Cosmology and Astroparticle Physics} {\bfseries 2019} no.~10, (Oct, 2019)
  031--031}.

\bibitem{musco:2019}
I.~Musco, ``Threshold for primordial black holes: Dependence on the shape of
  the cosmological perturbations'',
  \href{http://dx.doi.org/10.1103/PhysRevD.100.123524}{{\em Phys. Rev. D}
  {\bfseries 100} (Dec, 2019) 123524}.

\bibitem{young:2019}
S.~Young, I.~Musco, and C.~T. Byrnes, ``Primordial black hole formation and
  abundance: contribution from the non-linear relation between the density and
  curvature perturbation'',
  \href{http://dx.doi.org/10.1088/1475-7516/2019/11/012}{{\em Journal of
  Cosmology and Astroparticle Physics} {\bfseries 2019} no.~11, (Nov, 2019)
  012--012}.

\bibitem{byrnes:2019}
C.~T. Byrnes, P.~S. Cole, and S.~P. Patil, ``Steepest growth of the power
  spectrum and primordial black holes'',
  \href{http://dx.doi.org/10.1088/1475-7516/2019/06/028}{{\em Journal of
  Cosmology and Astroparticle Physics} {\bfseries 2019} no.~06, (Jun, 2019)
  028--028}.

\bibitem{satopolito:2019}
G.~Sato-Polito, E.~D. Kovetz, and M.~Kamionkowski, ``Constraints on the
  primordial curvature power spectrum from primordial black holes'',
  \href{http://dx.doi.org/10.1103/PhysRevD.100.063521}{{\em Phys. Rev. D}
  {\bfseries 100} (Sep, 2019) 063521}.
  \url{https://link.aps.org/doi/10.1103/PhysRevD.100.063521}.

\bibitem{Munoz:2017}
J.~B. Mu{\~{n}}oz, E.~D. Kovetz, A.~Raccanelli, M.~Kamionkowski, and J.~Silk,
  ``Towards a measurement of the spectral runnings'',
  \href{http://dx.doi.org/10.1088/1475-7516/2017/05/032}{{\em Journal of
  Cosmology and Astroparticle Physics} {\bfseries 2017} no.~05, (May, 2017)
  032--032}.

\bibitem{Motohashi:2017}
H.~Motohashi and W.~Hu, ``Primordial black holes and slow-roll violation'',
  \href{http://dx.doi.org/10.1103/PhysRevD.96.063503}{{\em Phys. Rev. D}
  {\bfseries 96} (Sep, 2017) 063503}.

\bibitem{gow:2020}
A.~D. Gow, C.~T. Byrnes, and A.~Hall, ``Primordial black holes from narrow
  peaks and the skew-lognormal distribution'', {\em arXiv preprint
  arXiv:2009.03204} (2020) .

\bibitem{musco:2021}
I.~Musco, V.~De~Luca, G.~Franciolini, and A.~Riotto, ``Threshold for primordial
  black holes. II. A simple analytic prescription'',
  \href{http://dx.doi.org/10.1103/PhysRevD.103.063538}{{\em Phys. Rev. D}
  {\bfseries 103} (Mar, 2021) 063538}.

\bibitem{byrnes:2021}
C.~T. Byrnes, E.~J. Copeland, and A.~M. Green, ``Primordial black holes as a
  tool for constraining non-Gaussianity'',
  \href{http://dx.doi.org/10.1103/PhysRevD.86.043512}{{\em Phys. Rev. D}
  {\bfseries 86} (Aug, 2012) 043512}.

\bibitem{inomata2021primordial}
K.~Inomata, E.~McDonough, and W.~Hu, ``Primordial Black Holes Arise When The
  Inflaton Falls'', {\em arXiv preprint arXiv:2104.03972} (2021) .

\bibitem{bird:pbhasdarkmatter}
S.~Bird, I.~Cholis, J.~B. Mu\~noz, Y.~Ali-Ha\"{\i}moud, M.~Kamionkowski, E.~D.
  Kovetz, A.~Raccanelli, and A.~G. Riess, ``Did LIGO Detect Dark Matter?'',
  \href{http://dx.doi.org/10.1103/PhysRevLett.116.201301}{{\em Phys. Rev.
  Lett.} {\bfseries 116} (May, 2016) 201301},
  \href{http://arxiv.org/abs/1603.00464}{{\ttfamily arXiv:1603.00464}}.

\bibitem{clesse:pbhmerging}
S.~Clesse and J.~Garc\'ia-Bellido, ``The clustering of massive Primordial Black
  Holes as Dark Matter: Measuring their mass distribution with advanced LIGO'',
  \href{http://dx.doi.org/https://doi.org/10.1016/j.dark.2016.10.002}{{\em
  Physics of the Dark Universe} {\bfseries 15} no.~Supplement C, (2017) 142 --
  147}, \href{http://arxiv.org/abs/1603.05234}{{\ttfamily arXiv:1603.05234}}.

\bibitem{brandt:ufdgconstraint}
T.~D. Brandt, ``Constraints on MACHO Dark Matter from Compact Stellar Systems
  in Ultra-faint Dwarf Galaxies'',
  \href{http://dx.doi.org/10.3847/2041-8205/824/2/L31}{{\em ApJ Letters}
  {\bfseries 824} no.~2, (2016) L31},
  \href{http://arxiv.org/abs/1605.03665}{{\ttfamily arXiv:1605.03665}}.

\bibitem{Munoz:2016}
J.~B. Mu\~noz, E.~D. Kovetz, L.~Dai, and M.~Kamionkowski, ``Lensing of Fast
  Radio Bursts as a Probe of Compact Dark Matter'',
  \href{http://dx.doi.org/10.1103/PhysRevLett.117.091301}{{\em Phys. Rev.
  Lett.} {\bfseries 117} (Aug, 2016) 091301}.

\bibitem{green:2016}
A.~M. Green, ``Microlensing and dynamical constraints on primordial black hole
  dark matter with an extended mass function'',
  \href{http://dx.doi.org/10.1103/PhysRevD.94.063530}{{\em Phys. Rev. D}
  {\bfseries 94} (Sep, 2016) 063530}.

\bibitem{zumalacarregui:supernovaconstraint}
M.~Zumalacarregui and U.~Seljak, ``{Limits on stellar-mass compact objects as
  dark matter from gravitational lensing of type Ia supernovae}'',
  \href{http://dx.doi.org/10.1103/PhysRevLett.121.141101}{{\em Phys. Rev.
  Lett.} {\bfseries 121} no.~14, (2018) 141101}.

\bibitem{murgia:pbh}
R.~Murgia, G.~Scelfo, M.~Viel, and A.~Raccanelli, ``Lyman-$\ensuremath{\alpha}$
  Forest Constraints on Primordial Black Holes as Dark Matter'',
  \href{http://dx.doi.org/10.1103/PhysRevLett.123.071102}{{\em Phys. Rev.
  Lett.} {\bfseries 123} (Aug, 2019) 071102}.

\bibitem{AliHaimoud:PBHmergerrate}
Y.~Ali-Ha\"{\i}moud, E.~D. Kovetz, and M.~Kamionkowski, ``Merger rate of
  primordial black-hole binaries'',
  \href{http://dx.doi.org/10.1103/PhysRevD.96.123523}{{\em Phys. Rev. D}
  {\bfseries 96} (Dec, 2017) 123523}.

\bibitem{deluca:nanograv}
V.~De~Luca, G.~Franciolini, and A.~Riotto, ``NANOGrav Data Hints at Primordial
  Black Holes as Dark Matter'',
  \href{http://dx.doi.org/10.1103/PhysRevLett.126.041303}{{\em Phys. Rev.
  Lett.} {\bfseries 126} (Jan, 2021) 041303}.

\bibitem{mukherjee:abhpbh}
S.~Mukherjee and J.~Silk, ``Can we distinguish astrophysical from primordial
  black holes via the stochastic gravitational wave background?'', {\em arXiv
  preprint arXiv:2105.11139} (2021) .

\bibitem{adamek:wimp_pbh}
J.~Adamek, C.~T. Byrnes, M.~Gosenca, and S.~Hotchkiss, ``WIMPs and stellar-mass
  primordial black holes are incompatible'',
  \href{http://dx.doi.org/10.1103/PhysRevD.100.023506}{{\em Phys. Rev. D}
  {\bfseries 100} (Jul, 2019) 023506}.

\bibitem{Carr:2016_rev}
B.~Carr, F.~K\"uhnel, and M.~Sandstad, ``Primordial black holes as dark
  matter'', \href{http://dx.doi.org/10.1103/PhysRevD.94.083504}{{\em Phys. Rev.
  D} {\bfseries 94} (Oct, 2016) 083504}.

\bibitem{Sasaki_2018}
M.~Sasaki, T.~Suyama, T.~Tanaka, and S.~Yokoyama, ``Primordial black
  holes{\textemdash}perspectives in gravitational wave astronomy'',
  \href{http://dx.doi.org/10.1088/1361-6382/aaa7b4}{{\em Classical and Quantum
  Gravity} {\bfseries 35} no.~6, (Feb, 2018) 063001},
  \href{http://arxiv.org/abs/1801.05235}{{\ttfamily arXiv:1801.05235}}.

\bibitem{Green:review}
A.~M. Green and B.~J. Kavanagh, ``Primordial black holes as a dark matter
  candidate'', \href{http://dx.doi.org/10.1088/1361-6471/abc534}{{\em Journal
  of Physics G: Nuclear and Particle Physics} {\bfseries 48} no.~4, (Feb, 2021)
  043001}.

\bibitem{Carr_Silk_2018}
B.~Carr and J.~Silk, ``{Primordial black holes as generators of cosmic
  structures}'', \href{http://dx.doi.org/10.1093/mnras/sty1204}{{\em MNRAS}
  {\bfseries 478} no.~3, (05, 2018) 3756--3775},
  \href{http://arxiv.org/abs/1801.00672}{{\ttfamily arXiv:1801.00672}}.

\bibitem{Carr20:review}
B.~Carr and F.~Kühnel, ``Primordial Black Holes as Dark Matter: Recent
  Developments'',
  \href{http://dx.doi.org/10.1146/annurev-nucl-050520-125911}{{\em Annual
  Review of Nuclear and Particle Science} {\bfseries 70} no.~1, (2020)
  355--394}.

\bibitem{Raidal19:pbh}
M.~Raidal, C.~Spethmann, V.~Vaskonen, and H.~Veermäe, ``Formation and
  evolution of primordial black hole binaries in the early universe'',
  \href{http://dx.doi.org/10.1088/1475-7516/2019/02/018}{{\em Journal of
  Cosmology and Astroparticle Physics} {\bfseries 2019} no.~02, (Feb, 2019)
  018--018}.

\end{thebibliography}\endgroup
\bibliographystyle{utcaps}

\end{document}